\documentclass[superscriptaddress, reprint, amsmath, amssymb,
aps,
pra,
floatfix]{revtex4-2}

\usepackage{mathptmx} 
\usepackage{blindtext}
\usepackage{xcolor} 
\usepackage{hyperref} 
\usepackage{array}
\hypersetup{
	colorlinks=true,
	linkcolor=blue,
	filecolor=magenta,
	urlcolor=magenta,
}

\usepackage{graphicx} 
\usepackage{subfigure} 
\usepackage{graphicx}
\usepackage{multirow}
\usepackage{orcidlink}
\usepackage{float}

\usepackage{amsmath}

\newcommand{\msi}[1]{\mbox{\scriptsize \textit{#1}}}

\begin{document}
\title{Input-Output Optics as a Causal Time Series Mapping: A Generative Machine Learning Solution}

\author{Abhijit Sen \orcidlink{0000-0003-2783-1763}}
\email{abhijit913@gmail.com}
\affiliation{Department of Physics and Engineering Physics, Tulane University,  New Orleans, Luisiana 70118, USA}

\author{Bikram Keshari Parida\orcidlink{0000-0003-1204-357X}}
\email{parida.bikram90.bkp@gmail.com}
\affiliation{Artificial Intelligence \& Image Processing Lab., Sun Moon University, Asan-si, South Korea}

\author{Kurt Jacobs\orcidlink{0000-0003-0828-6421}}
\email{dr.kurt.jacobs@gmail.com}
\affiliation{United States DEVCOM Army Research Laboratory, Adelphi, Maryland 20783, USA}
\affiliation{Department of Physics, University of Massachusetts at Boston, Boston, Massachusetts 02125, USA}

\author{Denys I. Bondar\orcidlink{0000-0002-3626-4804}}\email{dbondar@tulane.edu}
\affiliation{Department of Physics and Engineering Physics, Tulane University,  New Orleans, Luisiana 70118, USA}

\begin{abstract}
The response of many-body quantum systems to an optical pulse can be extremely challenging to model. Here we explore the use of neural networks, both traditional and generative, to learn and thus simulate the response of such a system from data. The quantum system can be viewed as performing a complex mapping from an input time-series (the optical pulse) to an output time-series (the systems response) which is often also an optical pulse. Using both the transverse and non-integrable Ising models as examples, we show that not only can temporal convolutional networks capture the input/output mapping generated by the system but can also be used to characterize the complexity of the mapping. This measure of complexity is provided by the size of the smallest latent space that is able to accurately model the mapping. We further find that a generative model, in particular a variational auto-encoder, significantly outperforms traditional auto-encoders at learning the complex response of many-body quantum systems. For the example that generated the most complex mapping, the variational auto-encoder produces outputs that have less than 10\% error for more than 90\% of inputs across our test data. 

\end{abstract}

\maketitle

\section{Introduction}

\begin{figure}
        \centering
        \includegraphics[width=\linewidth]{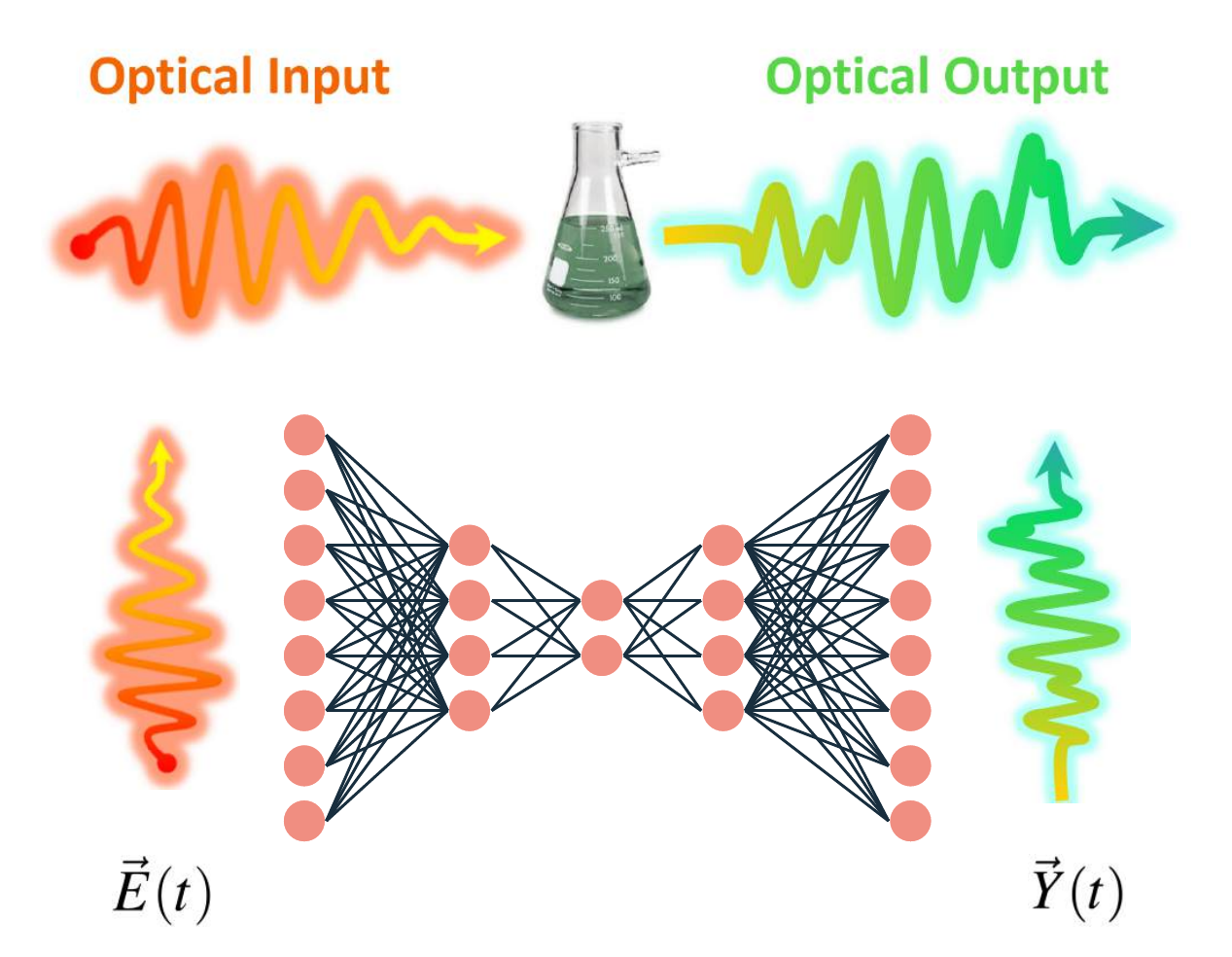}
        \caption{Optics as Input-Output Problem: The input (red pulse) gets transformed to an output (green pulse) after undergoing interaction with the substance. Upon training a neural network using the input-output data, the neural network replaces the substance and can be used for prediction of output pulse given an unseen input pulse.}
        \label{fig:first}
\end{figure}

Optics studies light and its interactions with matter. As depicted in Fig~\ref{fig:first}, the study comprises three main components: a quantum system, the incident pulse of light (optical input), and the system's response (optical output) to the incident pulse. The system's response usually varies depending on the input signal. The type of input signal can determine whether the dynamical system behaves linearly, non-linearly, or chaotically (a specific type of non-linear behavior characterized by extreme sensitivity) \cite{Strogatz2018}. If the input is weak, we have what is called linear optics. If the input intensity increases, we have a non-linear response/output signal, and the output can be expressed as a perturbation series expansion of the input~\cite{Boyd2008-ri}. As the intensity of the input field increases further, the system exhibits even more pronounced non-linearities requiring non-pertrubative treatment~\cite{amini_symphony_2019}, one of the most significant being \emph{High Harmonic Generation} (HHG). This strong-field laser–matter interaction resulting in HHG has attracted a lot of attention, and its discovery has been recognized with the 2023 Nobel Prize in Physics. HHG has been extensively studied in various media, including gases, semiconductors, nano-structures, metal, strongly
correlated materials, and solids \cite{hhgg1,hhgg2,hhgg3,hhgg4,hhgg5,hhgg6,hhgs1,hhgs2,hhgs3,hhgs4,hhgs5,hhgs6,hhgs7,hhgs1,hhgd2,hhgd4,hhgd5,hhgsc1,hhgsc2,hhgsc3,hhgsc4, korobenko_high-harmonic_2021}.  More recently, there is mounting interest in HHG from quantum spin systems due to their potential applications in probing spin dynamics and developing new laser sources in the THz regime \cite{Takayoshi2019, hhgqs1, hhgqs2, klimkin_coherent_2023}. 


The optical output (Fig~\ref{fig:first}) is obtained theoretically  by solving the time-dependent Schrödinger equation. Solving this equation rapidly becomes prohibitively expensive, as finding its solution is exponentially difficult with respect to the size of the quantum system with which the input pulse interacts. The complexity increases further if the quantum system is undergoing a phase transition. 

Light-matter interactions during phase transitions have been observed experimentally~\cite{alcala_high-harmonic_2022}. Understanding phase transitions is important for various reasons~\cite{pt1,pt2, pt3,pt4}. For example, in quantum computing, the transverse field Ising model is significant for quantum annealing, where phase transitions can influence the performance and efficiency of quantum algorithms \cite{Johnson2011}. The transverse Ising model is well known to exhibit a quantum phase transition and is analytically solvable \cite{phase2}. However, an Ising model with mixed transverse and longitudinal magnetic fields can only be solved numerically ~\cite{phase1}. In such regimes, small changes in system parameters can lead to large-scale changes in the system's dynamics, requiring extremely precise calculations that are challenging to achieve by numerically solving the Schrödinger equation. Alternative methods are needed to accurately model the behavior of such systems.

In this paper, we propose a novel data-driven methodology to model the input $\to$ output dynamics of optical systems using machine learning. We utilize \emph{Temporal Convolutional Networks} (TCNs) to capture the intricate interactions in many-body quantum systems by formulating optics as an input-output time series problem. While the presented methodology is generally applicable across different phenomena, we demonstrate our approach using the quantum Ising model as a test case. Additionally, we show how distinct physical regimes, including phase transitions, can be captured with machine learning models (see Fig.~\ref{fig:first}), tailored to the complexity of each regime.

The TCN is a recent addition to the toolbox of time series analysis.  TCNs use temporal convolution operations to enforce causality. By definition, the output of a convolution at  each time step is solely dependent on the past and not influenced by the future \cite{TCN8}. Further, the temporal dilation in a TCN helps in capturing long-range temporal patterns \cite{TCN1}. The efficacy of TCNs has led to their increasing appeal across multiple domains. Though more recent than recurrent neural networks (RNNs) and standard convolutional neural networks (CNNs), TCNs have already produced several noteworthy studies \cite{TCN1,TCN2,TCN1,TCN4,TCN5, TCN6, TCN7}. In Ref.~\cite{TCN7}, a study compared a TCN to a Long Short-Term Memory (LSTM) neural network for weather prediction, showing that the TCN demonstrated superior performance in time-series data forecasting. 

\begin{figure}
        \centering
        \includegraphics[width=\linewidth]{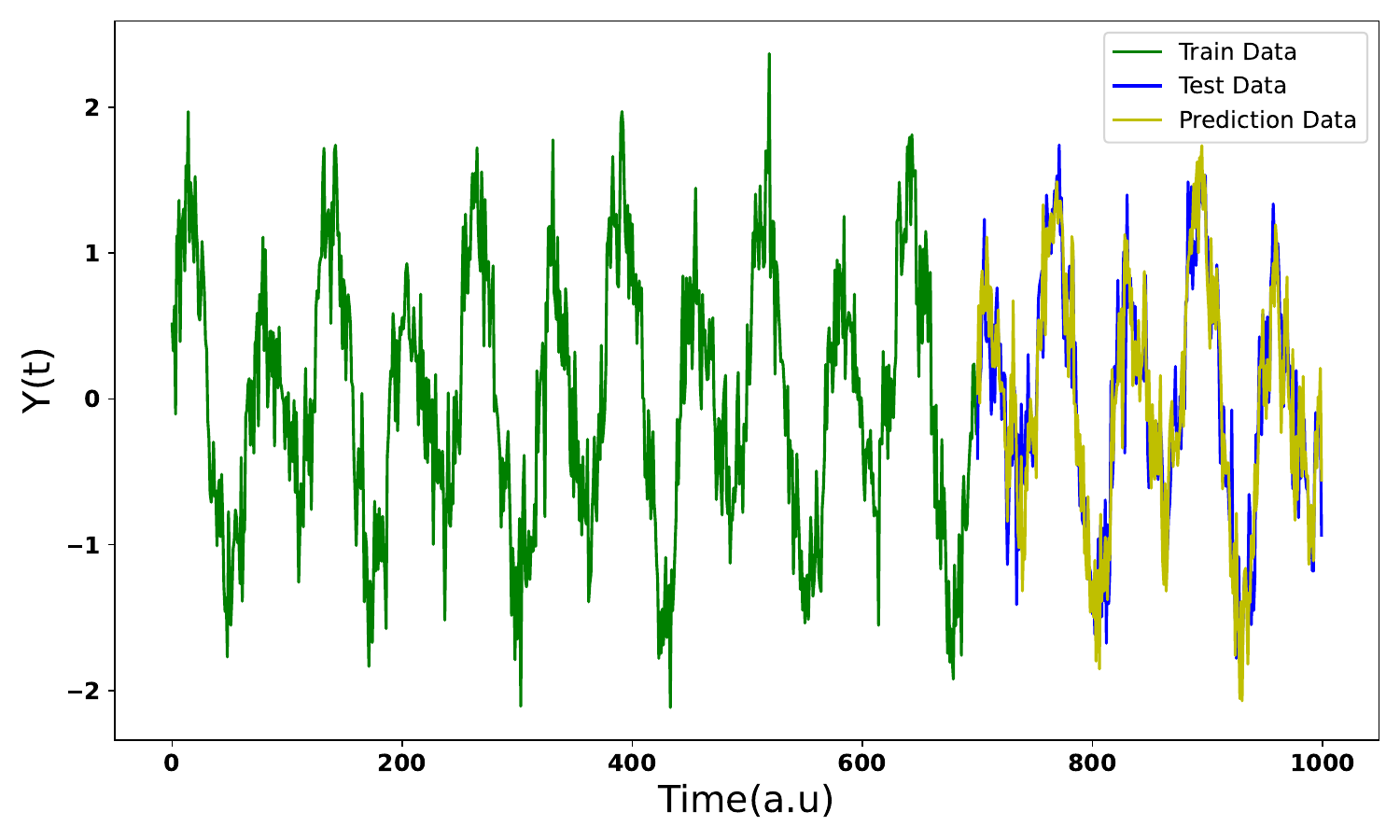}
        \caption{Schematic picture of time series modelling using TCN/LSTM: $Y(t)$ is the optical output as a function of time. Atomic units (a.u.) are used throughout. The blue portion depicts the historical data that serves as the training data for the model, the red portion depicts the test data or the unseen data for the model and the green curve depicts the model's predicted data whose closeness to the test data is a measure of the performance of the model.   }
        \label{fig:second}
\end{figure}

A conventional application of machine learning in time-series analysis is forecasting~\cite{TCN2, hochreiter1997long}. A notable example is Ref.~\cite{Yan2022}, where using a historical record (e.g., the training data shown in Fig.~\ref{fig:second}) for a time-domain output $Y(t)$, a model was constructed to predict $Y(t)$ for future times (see Fig.~\ref{fig:second}). It is important to note that this modeling does not utilize information across different input pulses. Hence, each time the incident laser pulse is changed, a new model needs to be trained. In the present work, we depart from this forecasting approach by training a TCN to learn the entire input $\to$ output relationship, employing a dataset of pairs of input and output pulses.

It is noteworthy that the inverse problem of characterizing output $\to$ input -- finding the optical input that produces the desired output -- has also been studied. Using quantum tracking control~\cite{ong_invertibility_1984, clark_quantum_1985, gross_inverse_1993, zhu_managing_1999, magann_singularity-free_2018, rothman_observable-preserving_2005, magann_quantum_2023}, it was proven~\cite{campos_how_2017, mccaul_driven_2020, mccaul_optical_2021} that by shaping the incident optical pulse, any desired output can be produced. This possibility implies that any two systems can be made spectrally identical, realizing an aspect of the alchemist's dream to make lead look like gold. We note that the output $\to$ input relation has previously been modeled using a CNN~\cite{lohani_dispersion_2019} by first converting the inputs into images.

The application of machine learning to light-matter interactions has been prolific. However, unlike our current work, the previous studies have not enforced causality.  These include applications in harmonic generation microscopy \cite{Shen2023}, nonlinear spectroscopy in solids \cite{Klimkin2023}, predicting characteristics of harmonic generation in plasmas using particle-in-cell simulations \cite{Mihailescu2016}, predicting HHG emission from molecules and inversely predicting molecular parameters from the corresponding spectra \cite{Lytova2023}, time series forecasting of nonlinear spectra \cite{Yan2022}, and predicting HHG from the spatially structured input fields~\cite{PablosMarn2023}. Additionally, deep neural networks have been employed in reconstruction of ultrashort pulses~\cite{ zahavy_deep_2018, brunner_deep_2022}, prediction of pulse properties generated by a free-electron laser~\cite{sanchez-gonzalez_accurate_2017}, and denoising measured photoelectron spectra~\cite{ kumar_giri_purifying_2020}.

Reformulating optics as an input-output time series analysis not only enables us to obtain quantitatively accurate models, but also allows us to import other techniques from the vast toolbox of time series analysis. For example, we introduce two new measures to quantify the complexity of optical dynamics. Traditionally, the complexity of a system’s dynamics has been assessed through the order of the perturbation series between the input and output \cite{mitra_identifying_2003, Boyd2008-ri, rey-de-castro_time-resolved_2013, maly_separating_2023}. The higher the order of the perturbation series, the more complex the dynamics.

\begin{figure}
        \centering
        \includegraphics[width=\linewidth]{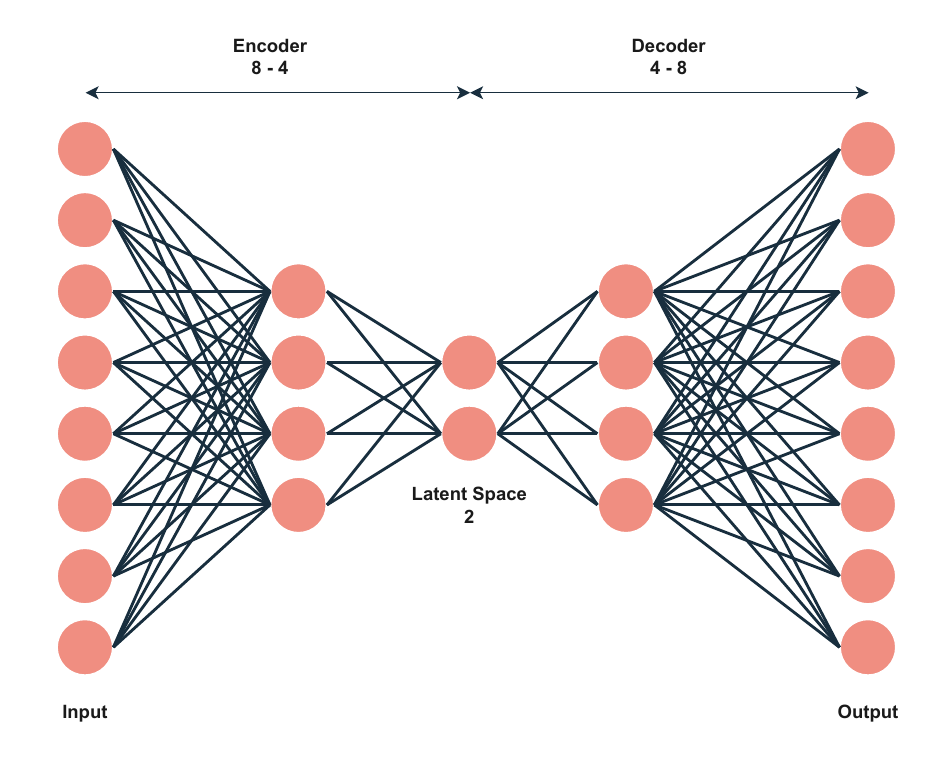}
        \caption{Deep Neural Network Autoencoder Architechture}
        \label{fig:three}
\end{figure}

The first method we introduce to quantify complexity, even in the non-perturbative regime, involves using the TCN to construct a model with an autoencoder architecture (Fig.~\ref{fig:three}). The purpose of the encoder is to compress the input data to its essential components (the ``latent space"),  and the necessary transformations are perform on this compressed data. The decoder then recovers the output from the transformed data. The dimension of the latent space -- the number of neurons it consists of -- can be taken as a measure of the complexity of the input-output relation. The larger the dimension, the more complex input $\to$ output transformation. It is noteworthy that this measure is reminiscent of the Kolmogorov complexity -- the length of the minimal program computing the desired output.

The second method for non-perturbative quantification of the complexity of the input-output dynamics employs amplitude-aware permutation entropy \cite{Ampentropy} of time series data, which provides a statistical measure of the system's disorder and complexity. To the best of our knowledge, neither the architecture of the machine learning model nor the amplitude-aware permutation entropy has been utilized  in optics.  

The rest of the paper is organized as follows: In Sec.~\ref{Section 2}, we comprehensively formulate the problem under investigation and provide details of the Ising system under consideration and various cases associated with it. Section~\ref{section3} is dedicated to presenting our results; we state details about the machine learning model architectures employed (Sec.~\ref{subA}), evaluation criteria (Sec.~\ref{subB}), illustrative examples (Sec.~\ref{subC})  that demonstrate the applicability of machine learning to the cases discussed earlier. In Sec.~\ref{subD}, we discuss the use of a more sophisticated generative machine learning architecture (Variational Autoencoder TCN) for better predictability. Further, we discuss in Sec.~\ref{subE} how the latent space of the architecture can be used to understand the complexity of the data. 

\section{Formulation of the problem}
\label{Section 2}
Due to recent theoretical and experimental interest in magnetic HHG using spin systems, we will study the input-output relation $h(t) \to Y(t)$ of the incident time-varying magnetic field $h(t)$ and magnetization $Y(t)$ for both the transverse and non-integrable Ising models. Note that time-varying magnetic fields, which are electromagnetic pulses with strong magnetic fields and low electric fields, can now be experimentally produced~\cite{Mukai2014}. 

The Hamiltonian of the transverse quantum  Ising model reads
\begin{equation}
\begin{split}
    \hat{H}_{\textit{t}} = & -\frac{1}{2}  h(t) \sum ^{N}_{n=1}\hat{\sigma} _{x}\left( n\right) -\frac{1}{2}\sum ^{N-1}_{n=1} J_{z}^{\left( n\right) }\hat{\sigma} _{z}\left( n\right) \hat{ \sigma} _{z}\left( n+1\right) \\ & - \frac{1}{2} J_{z}^{\left( N\right) }\hat{\sigma} _{z}\left( N\right) \hat{ \sigma} _{z}\left( 1\right),     
\end{split}
\label{T}
\end{equation}
where $\hat{\sigma} _{x}\left( n\right)$ and $\hat{\sigma} _{z}\left( n\right)$ are the $x$ and $z$ Pauli matrix acting on spin $n$, $J_{z}^{\left( n\right) }$ denotes the coupling energy of neighboring spins and  $h(t)$ is the input magnetic field. If $h(t)=0$ and all the couplings $J_{z}^{\left( n\right) }$ are positive, we recover the longitudinal Ising Hamiltonian with a ferromagnetic degenerate ground state. If the external magnetic field $h(t)$ is sufficiently strong, a phase transition from a ferromagnetic to a paramagnetic state takes place.

The non-integrable Ising model, which is  an Ising chain with  transverse ($h_1$) and longitudinal ($h_z$) constant magnetic fields, has the Hamiltonian of the form
\begin{equation}
\begin{split}
\hat{H}_{\textit{non-int}} = & -\frac{1}{2} (h_{1}+h(t)) \sum_{n=1}^{N} \hat{\sigma}_x(n) - \frac{1}{2} h_{z} \sum_{n=1}^{N} \hat{\sigma}_z(n) \\
& - \frac{1}{2} \sum_{n=1}^{N-1}  J_{z}^{(n)} \hat{\sigma}_z(n) \hat{\sigma}_z(n+1) - \frac{1}{2} J_{z}^{\left( N\right) }\hat{\sigma} _{z}\left( N\right) \hat{ \sigma} _{z}\left( 1\right). 
\label{NI}
\end{split}
\end{equation}

Quantum dynamics generated by Hamiltonians  \eqref{T} and  \eqref{NI} is rich due to  entanglement and phase transitions \cite{Chang2010,Kim2014}. We tune the parameters $h_{1}, h_{z},J_{z}^{\left( n\right) } $ in Eq \eqref{NI} to work in the phase transition regime.

For both the systems, we take the magnetization in the $x$-direction as an optical output
\begin{align}\label{EqResponseDer}
    Y(t) = \sum_{n=1}^{N} \left\langle \hat{\sigma}_x(n) \right\rangle,
\end{align}
where $ \left\langle \quad \right\rangle$ denotes the averaging over the time-dependent wave function.

To analyze the input~$\to$~output relation, $h(t) \to Y(t) $, via time-series analysis, we generate data sets by solving the Schr\"odinger equation with Hamiltonians~\eqref{T} and \eqref{NI} for $N=10$. (For this we employ the QuTiP library~\cite{qutip1, qutip2}.) We vary the site-to-site interaction strengths across the sites as $ \left( J_{z}^{(1)},\ldots, J_{z}^{(9)}\right) = ( 0.784 , 0.785, 0.787, 0.789, 0.791, 0.792, 0.794, 0.796, 0.798 ) $, which further adds to the complexity of the system response.

When performing numerical calculations, time is discretized into equally spaced small intervals of $\delta t$. Hence, instead of the continuous output signal $Y(t)$, we obtain the $T$-dimensional vector $\mathbf{Y}$ with components $Y_k = Y((k-1)\delta t)$, where $k=1, \ldots, T$. Thus, the problem of analyzing the input~$\to$~output relation, $h(t) \to Y(t)$, is equivalent to a vector~$\to$~vector transformation \(\mathbf{h} \to \mathbf{Y}\), where \(\mathbf{h}\) is the time-discretized input field with components \(h_k = h((k-1)\delta t)\), \(k=1, \ldots, T\).

For different datasets of samples of the input~$\to$~output pairs \(\{ \mathbf{h}^{(i)} \to \mathbf{Y}^{(i)} \}_i\), where \(i\) labels a sample, we build different TCN models of the time series transformation \(\mathbf{h} \to \mathbf{Y}\). In particular, we study the following three distinct cases in the subsequent sections:
\begin{itemize}
    \item[\textbf{Case 1:}]  Modeling the input~$\to$~output relation for the transverse Ising model~\eqref{T} based on a pre-calculated data set $\mathcal{D}(A = 1)$ of input-output pairs of time series, 
    \begin{align}
         \mathcal{D}(A) \equiv& \left\{  \mathbf{h}^{(i)} \to \mathbf{Y}^{(i)} \right\}_{i=1}^n \notag\\
       &\text{with } h^{(i)}_k = A \sin\left(\omega^{(i)} k \delta t\right),
        \label{caseA}
    \end{align}
    where $h^{(i)}_k$ is the $k$'th component of the $i$'th sample input vector $ \mathbf{h}^{(i)}$. The outputs are induced by  monochromatic inputs of different frequencies but identical amplitude $A$. To generate the data set the Sch\"odinger equation is solved $n$ times with  incident fields $h^{(i)}(t) = A \sin(\omega^{(i)} t)$ to obtain the corresponding responses $Y^{(i)}(t)$ via Eq.~\eqref{EqResponseDer}.

    \item[\textbf{Case 2:}]  Modeling the input~$\to$~output relation for the transverse Ising model~\eqref{T} from dataset $\mathcal{D}(A = 10)$ [Eq.~\eqref{caseA}], i.e., for strong driving field.

    \item[\textbf{Case 3:}]  Modeling the input~$\to$~output relation for the non-integrable Ising model~\eqref{NI} based on dataset $\mathcal{D}(A = 1.5)$ [Eq.~\eqref{caseA}], with the  parameters of Hamiltonian~\eqref{NI}  set to $h_{1} = -0.8\times1.05/2$, $h_{z} = 0.8\times0.5/2$ (see \cite{Kim2014}). 
    
    \item[\textbf{Case 4:}]  Modeling the input~$\to$~output relation for the non-integrable Ising model~\eqref{NI} based on dataset $\mathcal{D}(A = 2.5)$ with the parameters $h_{1} = -0.8\times1.05$, $h_{z} = 0.8\times0.5$. We note that  \textbf{Case 4} is more chaotic than \textbf{Case 3}.

    \item[\textbf{Case 5:}] Modeling the input~$\to$~output relation for the transverse Ising model~\eqref{T} based on the data set $\mathcal{D}(A_1, A_m)$ of inputs of varied intensity and frequencies,
\begin{align}
    \mathcal{D}(A_1, A_m) \equiv& \left\{ \mathbf{h}^{(i,j)} \to \mathbf{Y}^{(i,j)} \right\}_{i=1, \, j=1}^{n, \, m}  \notag\\
    & \text{with } h^{(i, j)}_k = A_j \sin\left(\omega^{(i)} k \delta t\right), \notag\\
     & A_j = A_1 + (j - 1) (A_m - A_1) / (m - 1),
    \label{caseB}
    \end{align}
    where $ h^{(i, j)}_k$ represents the $k$'th component of the $i$'th sample of input vector corresponding to amplitude $A_{j}$. Also, note that $A_1$ and $A_m$ denote the smallest and largest amplitudes, respectively. To obtain this data set, the Sch\"odinger equation must be solved $n \times m$ times. 
\end{itemize} 

Datasets~\eqref{caseA} and \eqref{caseB} are quite peculiar from the  point of view of ML applications. 
Dataset~\eqref{caseA} consists of two 2D arrays: $\{ \mathbf{h}^{(i)} \}_{i=1}^n$ storing  input pulses, $\{ \mathbf{Y}^{(i)} \}_{i=1}^n$ storing the induced output. Each of these arrays has a shape of $(n, T)$, where rows correspond to frequency and column to time. Correspondingly, dataset~\eqref{caseB} is a made of two 3D arrays of shape $(m, n, T)$, where the first dimension corresponds to the varied amplitude, the second to frequency, and the third to time. 

The studied transformation $\mathbf{h} \to \mathbf{Y}$ is in fact a multi-input and multi-output problem. The number of input and output features is $T$ -- the dimension of vectors $\mathbf{h}$ and $\mathbf{Y}$, thus leading to a high-dimensional dataset. When dealing with high-dimensional data,  the curse of dimensionality~\cite{Hughes1968} is encountered. To deal with this issue we increase the number of rows (that is,  increase the number of frequency values). Another approach to addressing the challenges posed by high-dimensional time series data is through data compression techniques \cite{Wang2008, Chiarot2023}. This method is preferable when the dataset is small. However, we employ the former approach here. 


\section{ Results} \label{section3}

The complete set of codes that reproduce all results presented below is available at \cite{Generative-ML-Quantum-System2025}.

\subsection{Model Architecture} \label{subA}
\begin{figure}
    \centering
    \includegraphics[width=\linewidth]{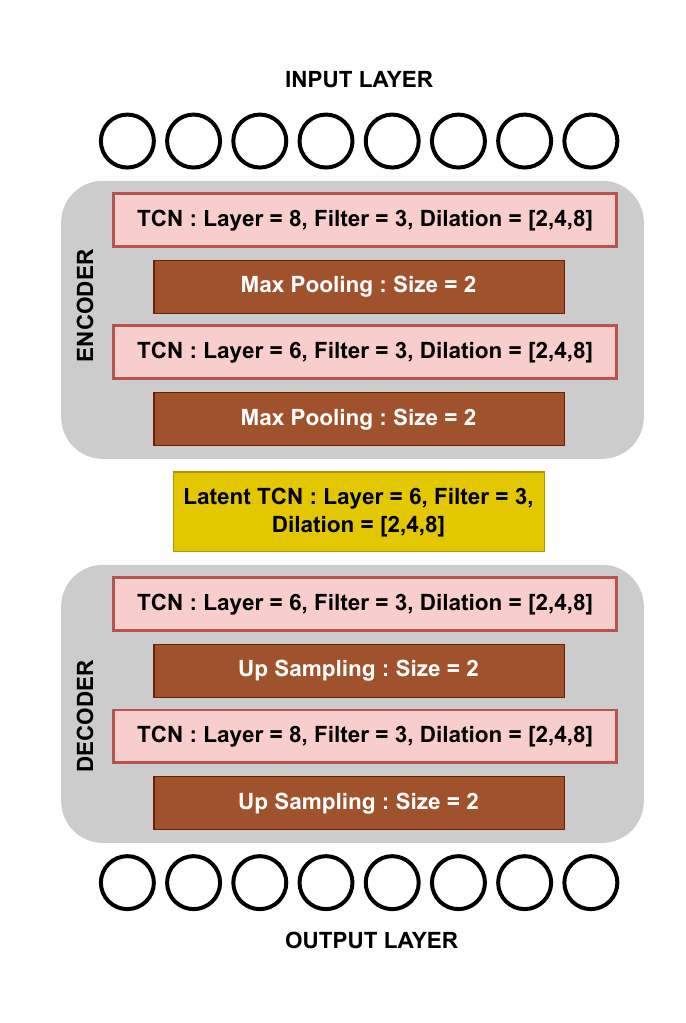}
    \caption{TCN Autoencoder Architecture}
    \label{fig:four}
\end{figure}

The choice of an effective model architecture is important in ML. In our analysis, we use autoencoder TCN architecture. To elucidate our approach, we first explore the encoder-bottleneck-decoder structure inherent to autoencoders within the context of deep neural networks. In Fig.~\ref{fig:three}, we present an example of a deep neural autoencoder architecture with a configuration of (8-4-2-4-8). This notation means that there are 8 neurons in the input layer followed by 4, 2, and 4 neurons in the subsequent three layers, respectively, and 8 neurons in the output layer. For simplicity, we will denote such a symmetrical architecture in short notation as (8-4-2), which represents the encoder part of the network, with the understanding that the decoder is its mirror image. The encoder takes an input and compresses it into a lower-dimensional representation, called a bottleneck or latent space. The latent space is a smaller vector representation of the input data. The decoder takes this latent representation and reconstructs the output \cite{VAE,auto2,auto3}. 
Similarly, we can have a TCN based autoencoder. An illustration of (8-6-6) autoencoder TCN framework is shown in Fig.~\ref{fig:four}.
We see the temporal 1D convolution layers along with max-pooling layers. In our auto-encoder based architecture, we use kernel/filter$=3$ and dilations ($2,4,8$). 

\subsection{Quantification of Predictive Quality} \label{subB}
We begin by introducing the problem of ML reproducibility. It is well-known that for a deep learning model, various factors can cause different outcomes in terms of accuracy in different runs, even with identical training data, identical algorithms, and identical networks \cite{reproduce1}. A simple way out of this problem is to run model training multiple times and record the accuracies of obtained models. If in multiple training runs, the accuracies do not fluctuate much (i.e., stay within $\sim 1\%$), we can safely conclude that the model is stable. In this paper, we find the \emph{minimum stable architecture} required to fit the input-output model.

We generate input-output datasets for \textbf{Cases 1 - 4} by setting  $n = 3700$ and $T = 512$ in Eq.~\eqref{caseA}. Therefore, for dataset pair $\{  \mathbf{h}^{(i)} \to \mathbf{Y}^{(i)} \}$, the $i$'th sample/row has input vector  $( h^{i}_1, h^{i}_2, \ldots, h^{i}_{512} )$ with the corresponding output vector  $ ( y^{i}_1, y^{i}_2, \ldots, y^{i}_{512} )$. After splitting the dataset into training $\left\{  \mathbf{h}^{(i)} \to \mathbf{Y}^{(i)} | i \in \text{ TrainSet} \right\} $ and testing  $\left\{  \mathbf{h}^{(i)} \to \mathbf{Y}^{(i)} | i \in \text{ TestSet} \right\}$ samples, we train a TCN model on the training sample. The obtained model can be used to produce the pair of data $\{  \mathbf{h}^{(i)} \to \mathbf{\widehat{Y}}^{(i)} | i \in \text{ TestSet} \}$, where $\mathbf{\widehat{Y}}^{(i)} = (\hat{y}^{i}_{1}, \hat{y}^{i}_{2}, \ldots, \hat{y}^{i}_{512})$ is the model's predicted output time series  for the given input time series  $\mathbf{h}^{(i)}$. Similarly, for \textbf{Case 5}, we generate the input-output dataset by setting $n = 3700$, $m = 7$ and $T = 512$ in Eq.~\eqref{caseB}, resulting a total of $n \times m$ samples. We specifically study the case of $A_1 = 0.5$, $A_m = 1.7$ with $m = 7$.

To quantify the model's predictive power, we calculate the R-squared value $R^2_i$ for entry  in the test set as follows:
\begin{align}\label{EqR2Def}
   R^2_i &= 1 - \frac{\sum_{k=1}^{T} (y^{i}_{k} - \hat{y}^{i}_{k})^2}{\sum_{k=1}^{T} (y^{i}_{k} - \bar{y}_{i})^2}, \quad i \in \text{ TestSet}, \\
    & \bar{y}_{i} = \frac{1}{T} \sum_{k=1}^{T} y^{i}_{k}. \notag
\end{align}

The model is considered to be stable if, across multiple training runs, more than 85\% of the R-squared values in the set $\{R^2_i | i \in \text{ TestSet} \}$ are above 0.85.

\subsection{Illustrations and Discussions} \label{subC}

\begin{figure}[t]
        \centering
        \includegraphics[width=\linewidth]{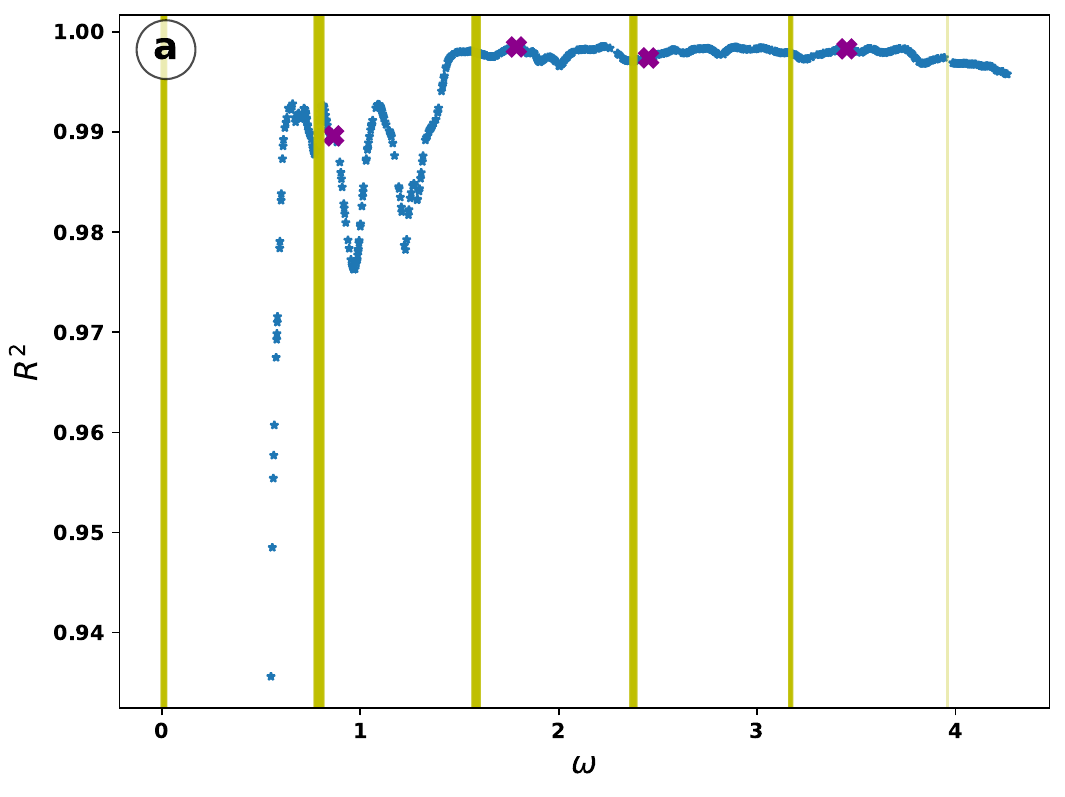}
        \includegraphics[width=\linewidth]{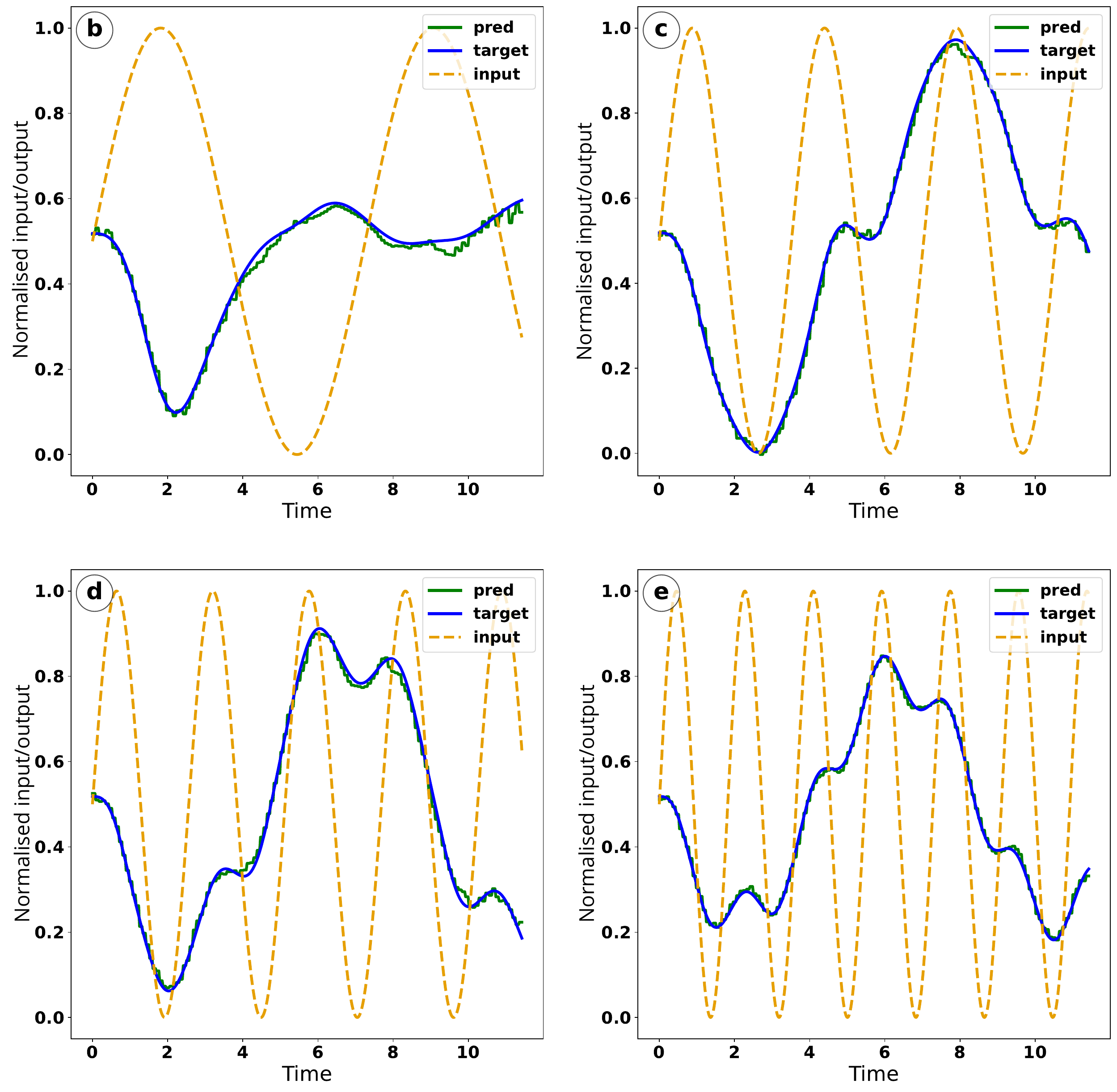}
        \caption{Performance illustration of a temporal convolutional network (TCN) model constructed for \textbf{Case 1} of Sec.~\ref{Section 2}. Plot (a) shows  the $R^{2}$ values vs the frequency $\omega$ for the test data set, i.e., $\{(\omega^{(i)}, R^2_i), \mid i  \in \text{TestSet}\}$ [Eqs.~\eqref{caseA} and \eqref{EqR2Def}]. The vertical green lines mark the transition frequencies of the time-independent Hamiltonian, demonstrating that the constructed time series model performs uniformly across the entire frequency range. The purple crosses mark the frequencies $\{\omega^{(i)}\} = \{0.86, 1.78, 2.45, 3.45\}$ of the optical inputs $\mathbf{h}^{(i)}$ [Eq.~\eqref{caseA}], labeled as \textbf{input} in subplots~(b)-(e). The   output time series $\mathbf{Y}^{(i)}$  [Eq.~\eqref{caseA}] (labeled as \textbf{target}) induced by the inputs $\mathbf{h}^{(i)}$ 
        and the outputs $\mathbf{\widehat{Y}}^{(i)}$ predicted by the time series model (\textbf{pred}) are also shown in subplots~(b)-(e).
        }\label{fig:five}
\end{figure}

The minimum autoencoder TCN architecture for \textbf{Case 1} in Sec. \ref{Section 2}  is (5-5-3) with 2082 trainable parameters. To evaluate the stability of the model architecture, we trained 10 different models of the same architecture. For each such model, above $90\%$ of  $\{R^2_i | i \in \text{ TestSet} \}$ lie above $0.98$, indicating stability of the architecture.  Fig.~\ref{fig:five} illustrates the performance of one such model.

In Fig.~\ref{fig:five}(a), we observe an oscillatory behavior for $R^2$ at low frequencies $\omega$. To determine whether this oscillation reveals any underlying physics, we compare it with the characteristic transition frequencies of the quantum system $\omega_{kl} = (E_k - E_l)/\hbar$, $E_k > E_l$, (recall that $\hbar = 1$ in atomic units), where $E_m$ are eigenenergies (i.e., eigenvalues) of the time-independent part of the Hamiltonian. Physically, $\omega_{kl}$ denotes the energies of photons required to drive transitions between energy levels $E_l$ and $E_k$. In Fig.~\ref{fig:five}(a), the vertical green lines mark the transition frequencies $\omega_{kl}$,  which cluster into bands. Since no correlation is observed between $R^2$ and the transition frequencies,   we conclude that the time-series model performs uniformly across the entire frequency range.

\begin{figure}[t]
        \centering
        \includegraphics[width=\linewidth]{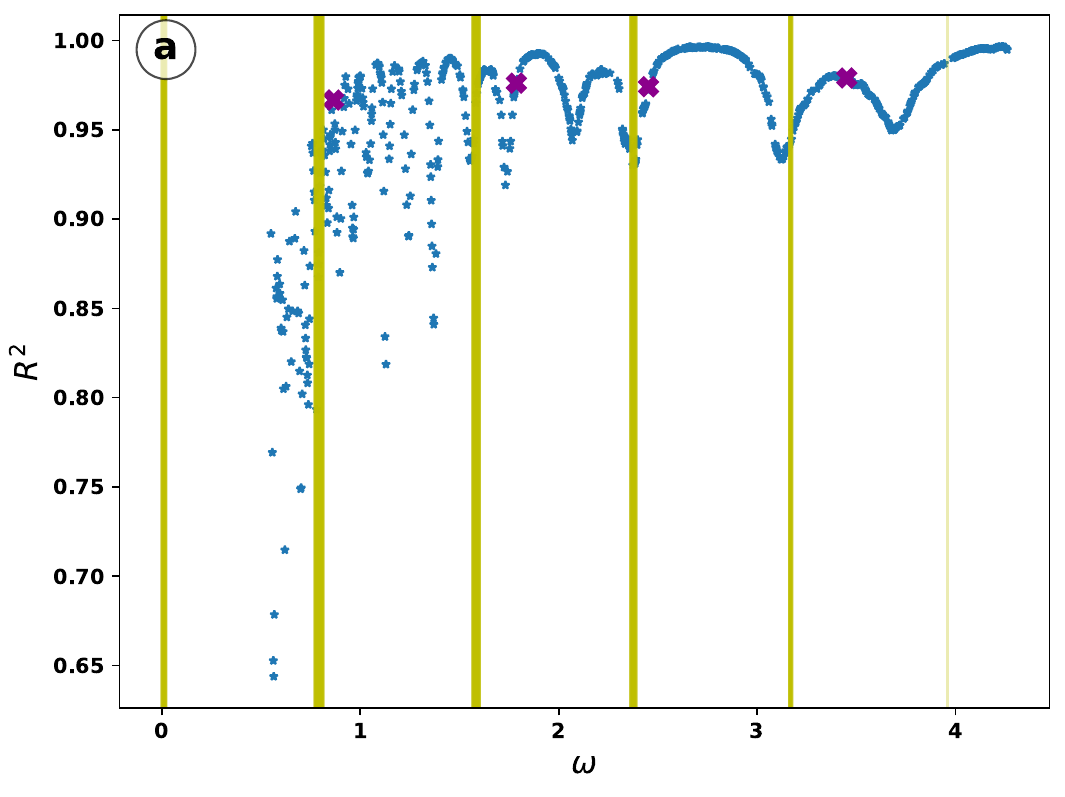}
        \includegraphics[width=\linewidth]{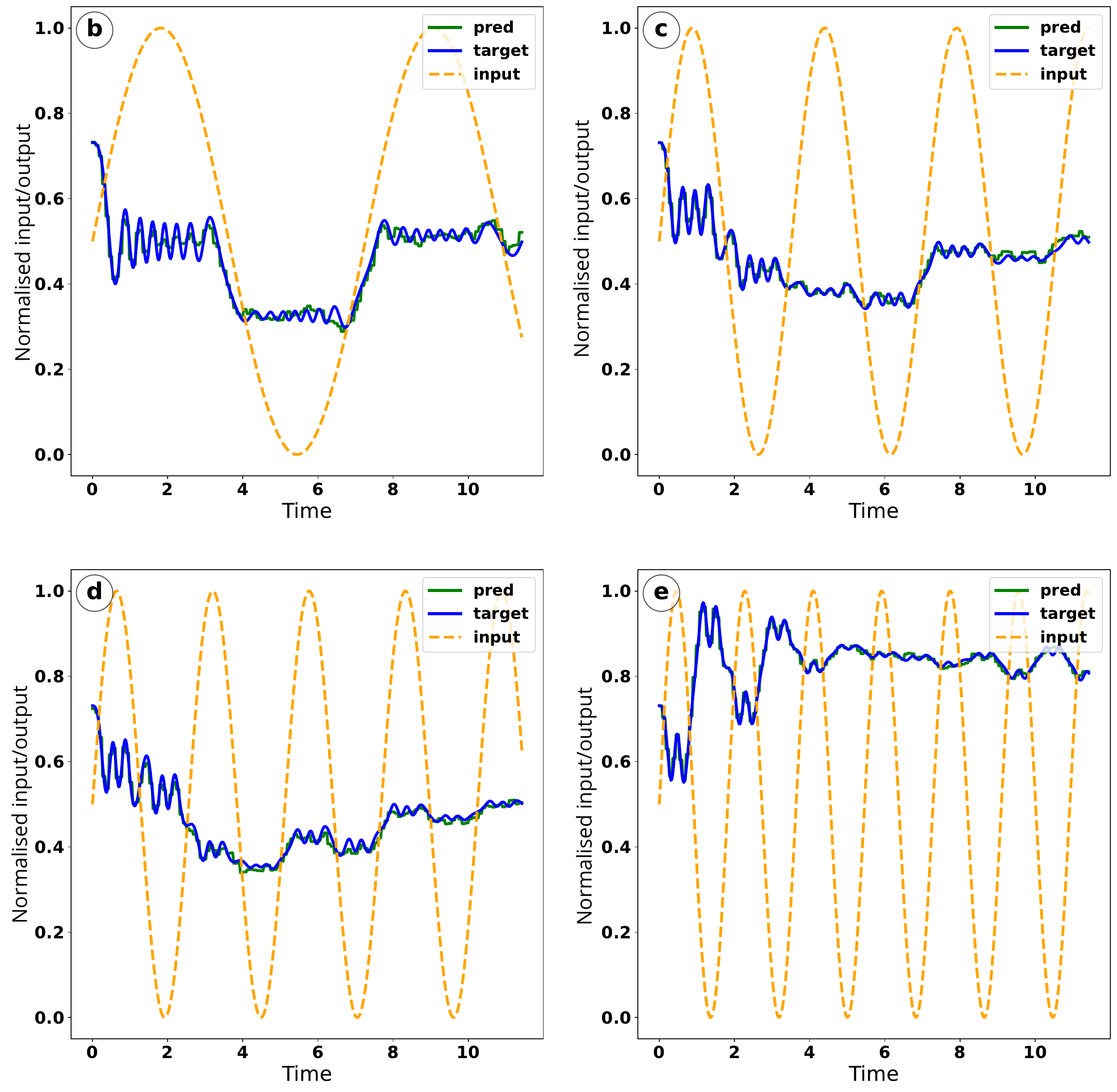}
        \caption{
        Performance illustration of a temporal convolutional network (TCN) model constructed for \textbf{Case 2} of Sec.~\ref{Section 2}. See the caption of Fig.~\ref{fig:five} for further details.}
        \label{fig:six}
\end{figure}

Similarly, in \textbf{Case 2} of Sec.~ \ref{Section 2}, the minimum stable autoencoder TCN architecture is (12-12-10-10) with 16127 trainable parameters. We again performed a stability assessment by training 10 models of the same architecture and found that for any  model, above $85\%$ of  $\{R^2_i | i \in \text{ TestSet} \}$ lie above are above $0.90$ (see Fig \ref{fig:six} for one such model). The need for a deeper architecture suggests that the dynamics in \textbf{Case 2} are more non-linear then in \textbf{Case 1}, resulting in a more intricate output dataset. An independent complexity analysis in the next section will confirm this. Furthermore, we also applied a variational TCN autoencoder (VAE) architecture to case 2 to see if this would provide a significantly better model. It did so and we present the results in in \ref{subD}.

\begin{figure}[t]
        \centering
        \includegraphics[width=\linewidth]{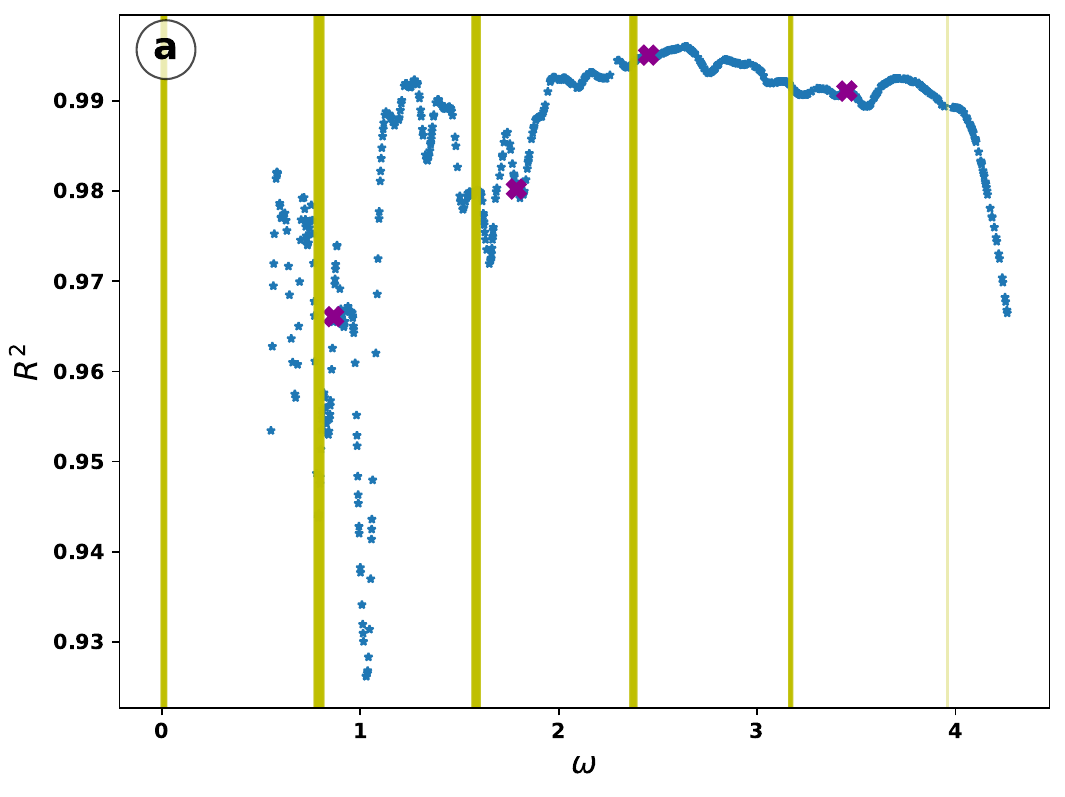}
        \includegraphics[width=\linewidth]{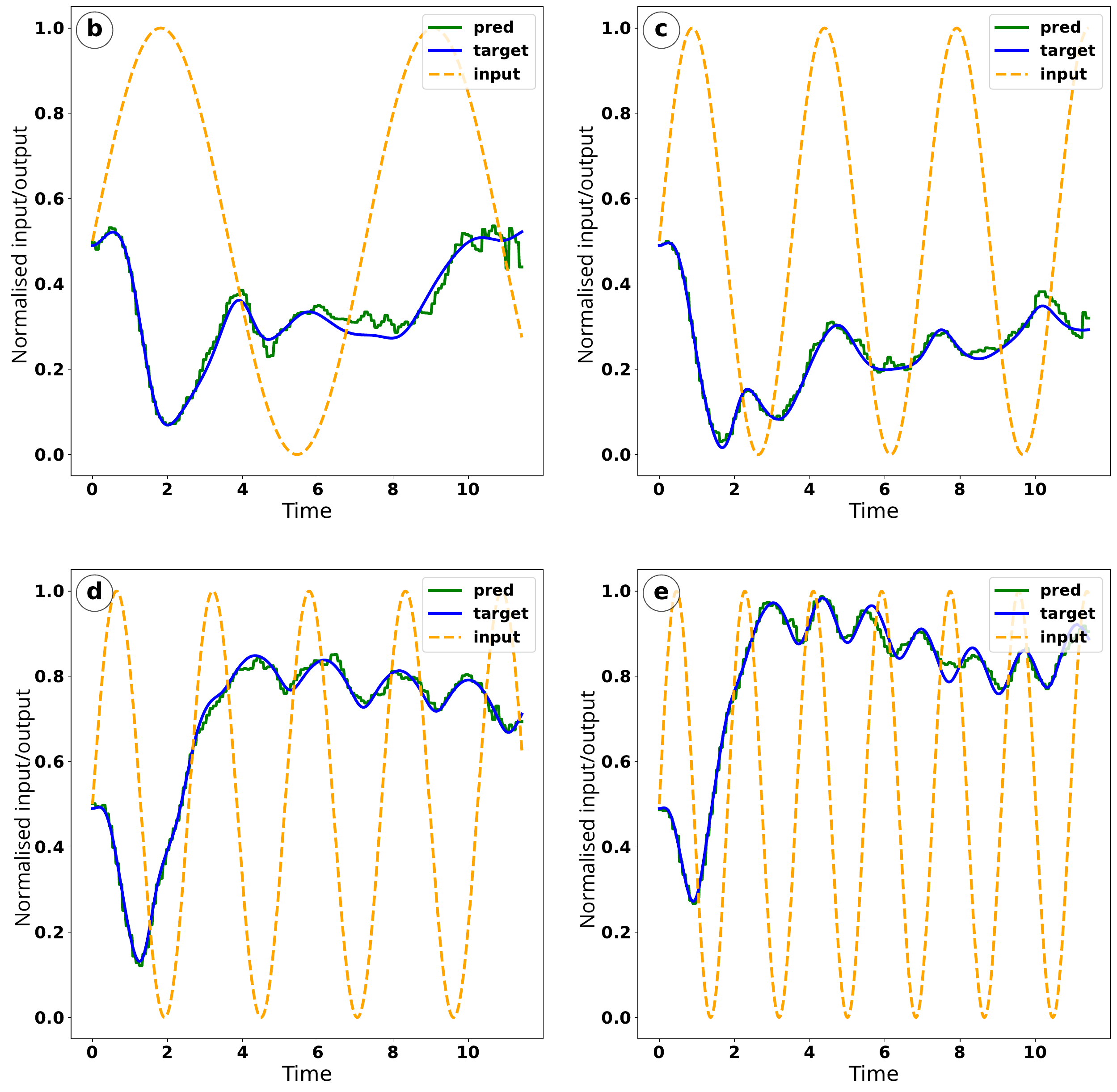}
        \caption{
        Performance illustration of a temporal convolutional network (TCN) model constructed for \textbf{Case 3} of Sec.~\ref{Section 2}. See the caption of Fig.~\ref{fig:five} for further details.}
        \label{fig:seven} 
\end{figure}

\begin{figure}[t]
        \centering
        \includegraphics[width=\linewidth]{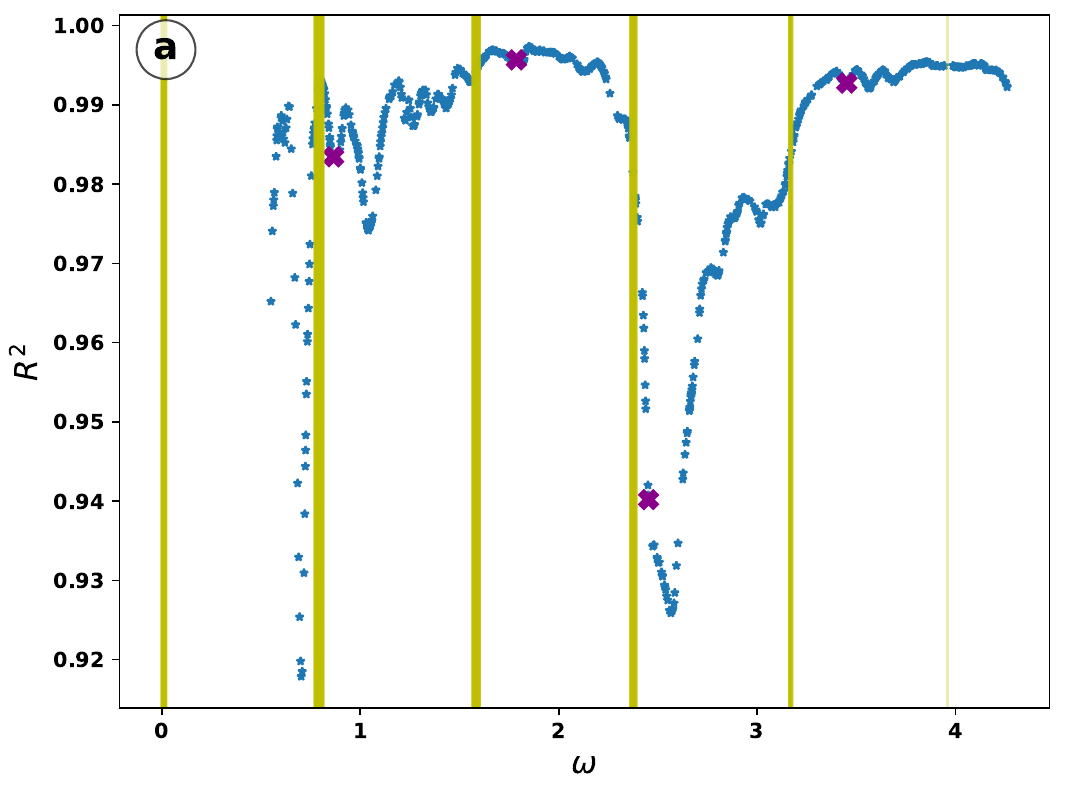}
        \includegraphics[width=\linewidth]{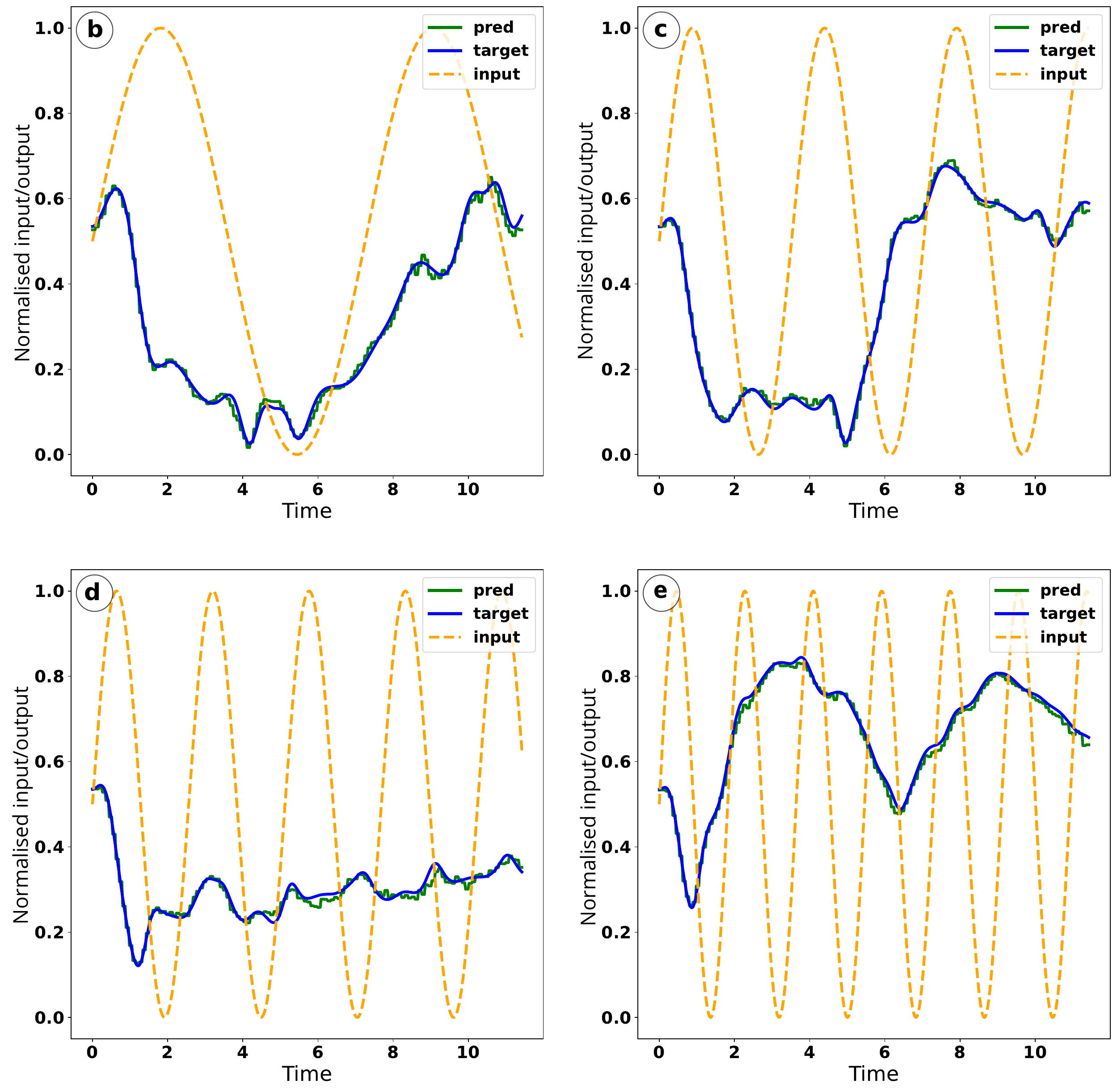}
        \caption{
        Performance illustration of a temporal convolutional network (TCN) model constructed for \textbf{Case 4} of Sec.~\ref{Section 2}. See the caption of Fig.~\ref{fig:five} for further details.}
        \label{fig:eight}
\end{figure}

In \textbf{Case 3} of Sec.~\ref{Section 2}, the minimal stable autoencoder TCN architecture is (5-5-4) with 2,234 trainable parameters. We trained 10 different models using this architecture and observed that, for each model, over $95\%$ of  \(\{R^2_i | i \in \text{ TestSet} \}\) consistently exceed 0.95  (see Fig \ref{fig:seven} illustrating one such model). In \textbf{Case 4} we observe a stable TCN architecture has a configuration of (8-8-6) with 5,439 trainable parameters. Similar to \textbf{Case 3}, we trained 10 models with this architecture. For each model, over $90\%$ of  \(\{R^2_i | i \in \text{ TestSet} \}\) also consistently exceed 0.95  (see Fig \ref{fig:eight} showing performance of one such model). 

In \textbf{Case 5}, we evaluated 10 distinct models with an (8-8-4) architecture with 3,221 parameters. Remarkably, all models achieved an R-squared value exceeding 0.95, with at least $88\%$ of  \(\{R^2_i | i \in \text{ TestSet} \}\) surpassing this threshold.

In all the cases treated, we ensured a similar train-test split of the data and kept the number of epochs the same. For all our training processes we used the swish activation function~\cite{swi}, the Adam optimizer with a learning rate of 0.001, and for the loss function we used the Huber loss~\cite{huber}. Additionally, we employed the technique of changing batch sizes~\cite{BATCH}. For all cases we used TCNs with dilation factors of 2, 4, and 8 to effectively capture long-range dependencies in the dataset. This approach allowed us to expand the receptive field of the convolutional layers. Further, in all plots, the input, output, and predicted values are scaled between 0 and 1 because the normalization facilitates a consistent comparison. The plots remain consistent and accurate after applying the inverse transform. 

We have summarized the training details and the results in Tables \ref{training} and \ref{tabular_result}, respectively.

\begin{table}
\centering
\begin{tabular}{|>{\raggedright\arraybackslash}p{3cm}|>{\raggedright\arraybackslash}p{5cm}|}
\hline
Model Type           & Temporal Convolutional Networks                                                             \\ \hline
Activation Function  & Swish                                                                                               \\ \hline
Loss Function      & Huber Loss ([Reference 87])                                                                         \\ \hline
Learning Rate       & 0.001                                                                                               \\ \hline
Optimizer         & Adam                                                                                                \\ \hline
Train-Test Split     & Similar split across all cases                                                                      \\ \hline
Epochs              & 250                                                                                                \\ \hline
Dilation Factors     & 2, 4, and 8 (used to capture long-range dependencies)                                               \\ \hline
\end{tabular}
\caption{Summary of Training Configurations}
\label{training}
\end{table}

We note that as the complexity of input-output data increases, it is important to examine the behavior of the latent space within the encoder-decoder model. The encoder and decoder collaboratively compress the data into a lower-dimensional representation known as the latent space. When the complexity of the input-output relationship increases, the dimensionality of the latent space must also increase. If the latent dimension is reduced beyond a necessary threshold, the compression quality deteriorates, leading to an insufficient capture of essential features in the latent space. Thus suggests that the size of the smallest latent space that is able to produce an accurate model for the input-output system is a good measure of the complexity of the system. Among the five cases we considered this measure suggests that \textbf{Case 2} is the most complex. In Sec.~\ref{subE}, we evaluate the complexity of the input-output transformations using a more traditional measure of the complexity of time-series. This provides additional evidence that the size of the latent space is a good measure of the system complexity. 


\begin{table*}
\centering

\begin{tabular}{|c|p{1.7cm}|p{1.45cm}|c|c|c|c|c|p{5cm}|}
\hline
\textbf{Cases} & \textbf{Hamiltonian} & \textbf{Amplitude}  & \textbf{Algorithm} & \textbf{Architecture} & \textbf{Parameters} & \textbf{Loss Function} & \textbf{Epochs} & \textbf{Result} \\ \hline
 1 &  Eq.~\eqref{T} & $A = 1$ & TCN & 5-5-3 & 2082 & Huber Loss & 250 &over $90\%$ of $\{R^2_i | i \in \text{ TestSet} \}$ consistently exceed  $0.98$ \\ \hline
 2 &  Eq.~\eqref{T} & $A = 10$ & TCN & 12-12-10-10 & 16127 & Huber Loss & 250& over $85\%$ of $\{R^2_i | i \in \text{ TestSet} \}$ consistently exceed  $0.90$ \\ \hline
 2 &  Eq.~\eqref{T} & $A = 10$ & VAE-TCN & 12-12-10-10 & 34127 & Huber Loss & 250&  over $90\%-95\%$ of $\{R^2_i | i \in \text{ TestSet} \}$ consistently exceed  $0.90$ \\ \hline
 3 &   Eq.~\eqref{NI} & $A = 1.5$ & TCN & 5-5-4 & 2234 & Huber Loss & 250 & over $95\%$ of \(\{R^2_i | i \in \text{ TestSet} \}\) consistently exceed $0.95$ \\ \hline
 4 &   Eq.~\eqref{NI} & $A = 2.5$ & TCN & 8-8-6 & 5439 & Huber Loss & 250 & over $90\%$ of \(\{R^2_i | i \in \text{ TestSet} \}\) consistently exceed $0.95$ \\ \hline
 5 &   Eq.~\eqref{NI} &$A_1 = 1$ $A_m = 1.7$ & TCN & 8-8-4 & 3221 & Huber Loss & 250 & over $88\%$ of $\{R^2_i | i \in \text{ TestSet} \}$ consistently exceed $0.95$ \\ \hline
\end{tabular}
\caption{Performance evaluation of time series algorithms for the cases listed in Sec.~\ref{Section 2}, studying different driven Ising systems. The results highlight the proportion of test set predictions achieving high R-squared values.}
\label{tabular_result}
\end{table*}

\subsection{Variational TCN Autoencoder} \label{subD}

Variational autoencoders (VAEs) \cite{9051780,VAE} are advanced versions of traditional autoencoders (AEs) with the ability to extract higher-level abstractions from input data. The key difference between VAEs and AEs lies in their approach to latent space representation. VAEs employ a probabilistic latent space, enabling them to capture a richer and more structured representation. In contrast, AEs rely on a deterministic latent space, which can be less organized and may limit their capacity to generalize. Since VAEs generate outputs from a learned probability density, they are typically regarded as \textit{generative} models.

In a VAE, each input is encoded into a mean and variance, which define a probability distribution (typically Gaussian). During training, the model samples from this learned distribution to generate latent variables. This sampling process allows VAEs to learn robust features. The latent space is regularized to follow a known distribution using the Kullback-Leibler divergence, incorporated into the loss function alongside the reconstruction loss. (Minimizing the reconstruction loss ensures that the model reconstructs the input data samples as accurately as possible, contributing to a robust latent space.)

\begin{figure}
        \centering
        \includegraphics[width=\linewidth]{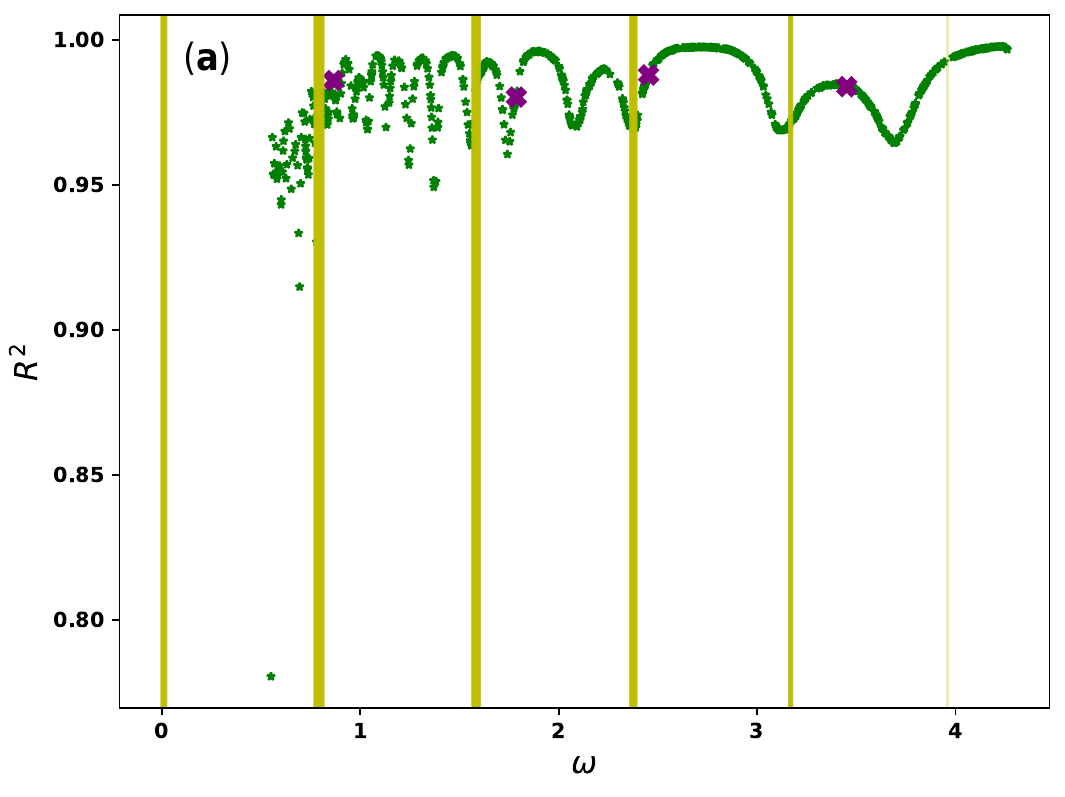}
        \includegraphics[width=\linewidth]{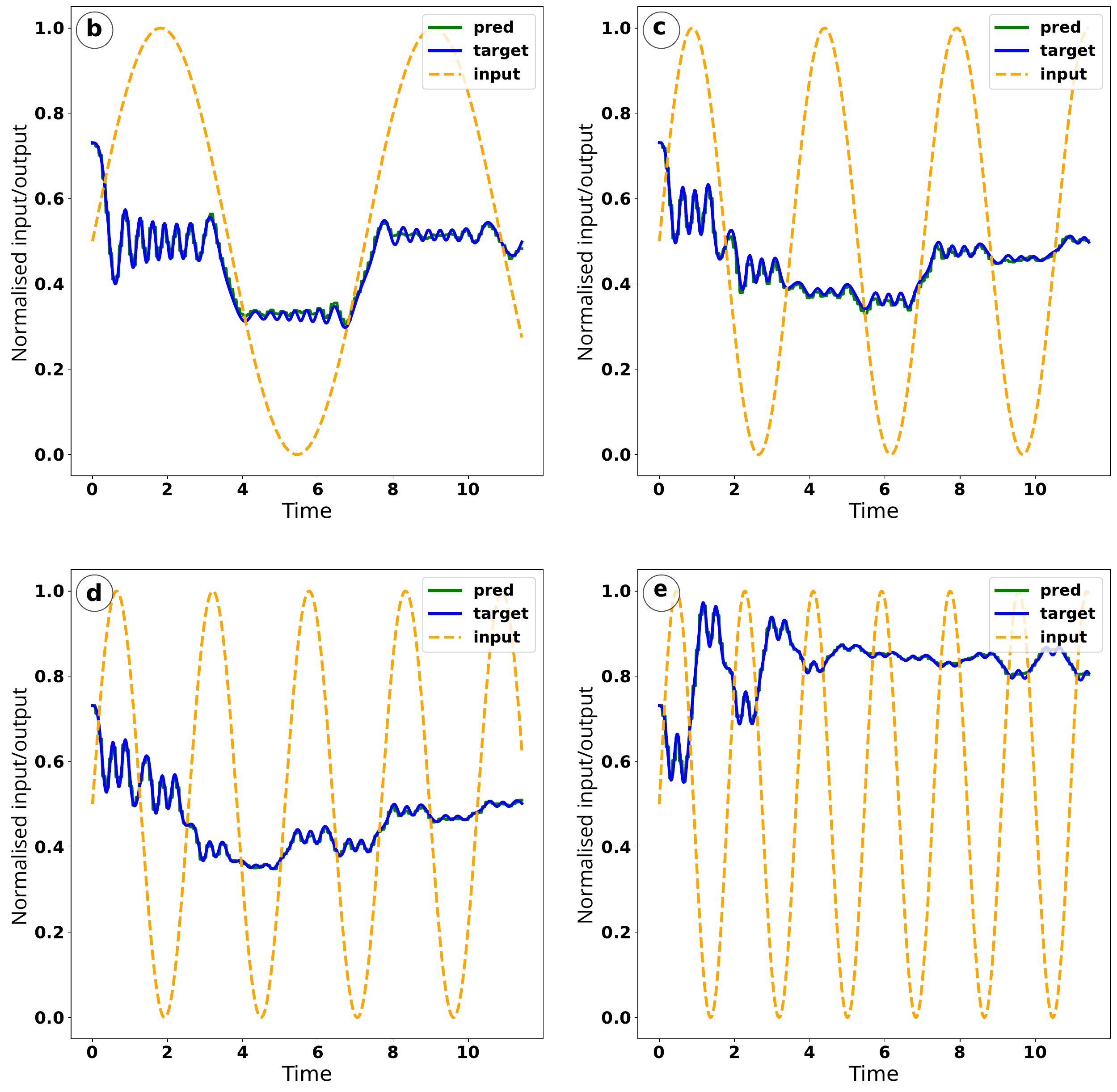}
        \caption{
        Performance illustration of a variational  temporal convolutional network autoencoder (VAE TCN) model constructed for \textbf{Case 2} of Sec.~\ref{Section 2}.
        See the caption of Fig.~\ref{fig:five} for further details.}
        \label{fig:ten}
\end{figure}

We implement a TCN VAE for \textbf{Case 2} of Sec.~\ref{Section 2} with a TCN architecture consistent with our previous design: (12-12-10-10), incorporating additional sampling layers. The number of parameters is 34,127, greater than the number of parameters in the traditional AE used in Sec.~\ref{subC}. We trained 10 different models with the (12-12-10-10) TCN VAE architecture and found that these models were able to reach a Huber Loss value of $\mathcal{O}(10^{-5})$ within a small number of epochs, significantly lower than $\mathcal{O}(10^{-4})$ -- the minimum loss achieved by AE. Furthermore, we found that over $90\% - 95\%$ of the R-squared values in \(\{R^2_i | i \in \text{ TestSet} \}\) also consistently exceeded $0.90$ as shown in Fig.~\ref{fig:ten}.

\subsection{Complexity Analysis}\label{subE}

Entropy provides a valuable metric for assessing the level of uncertainty and irregularity within a signal or system. Among the various entropy measures, the permutation entropy is known for its robustness in analyzing the complexity of time series data \cite{permutation1,permutation2}. However, since HHG dynamics strongly depends on the amplitude of the incident pulse, we use the amplitude-aware permutation entropy by deploying the package Entropy Hub \cite{hub}, which overcomes limitations of the traditional permutation entropy \cite{Ampentropy}. To enable a standardized comparison of complexity across different conditions, we employ Pnorm, or Mean Normalized Permutation Entropy, which normalizes the permutation entropy to a fixed range, providing an averaged measure of time series complexity.

We now describe how we employ the amplitude-aware permutation entropy to obtain a measure of the overall complexity of the output dataset generated for any amplitude $A$. In data set~\eqref{caseB} for quantum system~\eqref{T}, we set $n=200$, $m = 19$ and $T = 512$. The minimum and maximum amplitudes taken are $A_{1} = 1$ and $A_{m} = 10$ respectively. For each amplitude $A_{j}$ from data set~\eqref{caseB}, we compute the mean normalized permutation probabilities (Pnorm) $M_{j}$ (with embedding dimension equals $4$) of the elements in set $C_{A_{j}} = \{ C_{i} \mid i = 1, 2, \ldots, n \}$. This set $C_{A_{j}}$ consists of $n$ values, $C_{i}$, each representing the Pnorm calculated for the $i^{\msi{th}}$ sample in the output dataset $Y^{(i,j)}$. Our results indicate a consistent increase in Pnorm values with increasing amplitude, reflecting a corresponding rise in the complexity of the output dataset. We also applied the same method for \textbf{Cases 3} and \textbf{4} of Sec.~\ref{Section 2}. We observed that the mean normalized permutation probabilities  for these amplitudes were higher, indicating increased complexity in the dynamics during phase transitions.

\begin{figure}
        \centering
        \includegraphics[width=\linewidth]{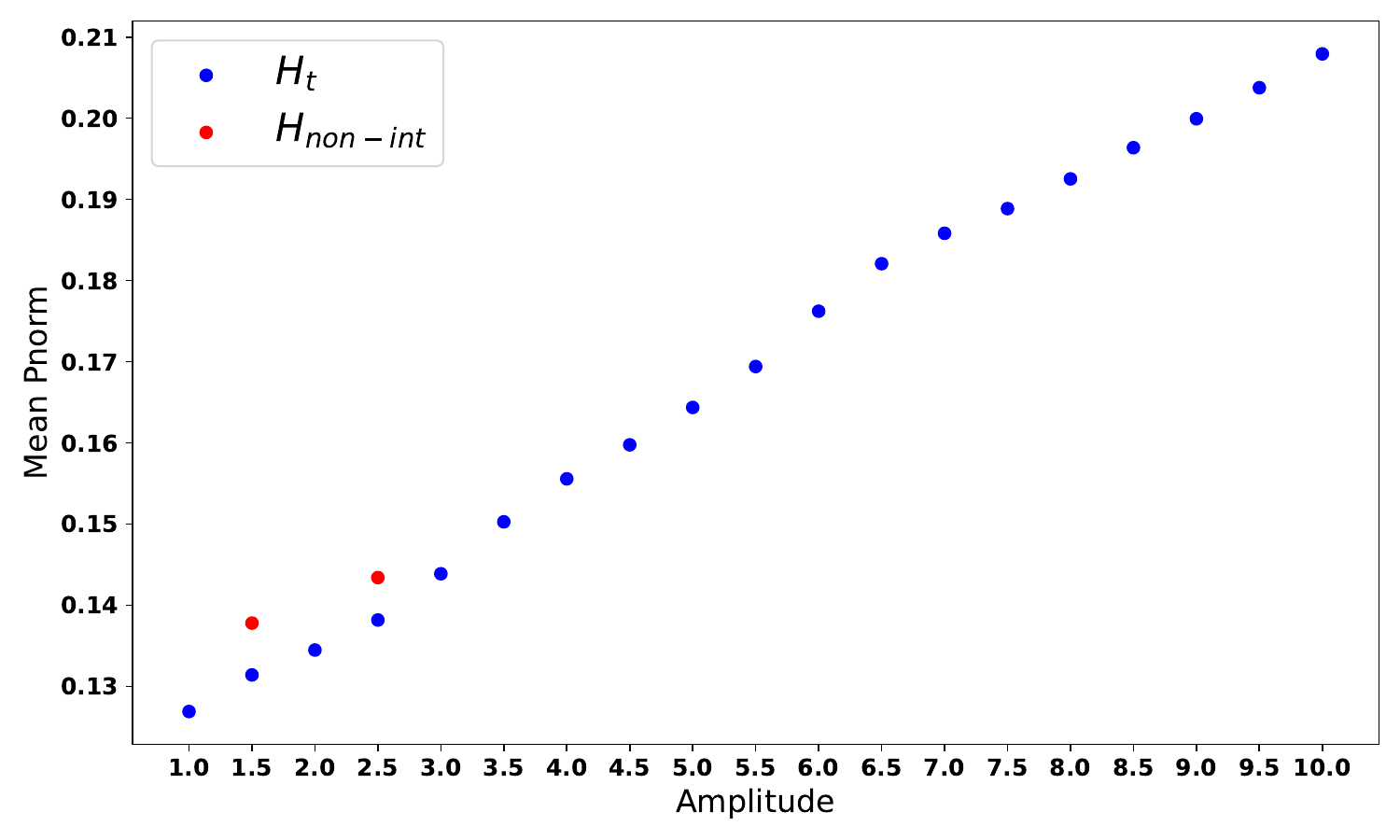}
        \caption{The mean normalized permutation probabilities (Pnorm) versus the amplitude of the input pulse. This plot shows that as the amplitude increases, the complexity of the optical output also increases. The blue dots represent the mean Pnorm values for \textbf{Case 5} of Sec.~\ref{Section 2}, characterizing the transverse Ising model~\eqref{T}. The two red dots correspond to the mean Pnorm values for \textbf{Cases 3} and \textbf{4}, respectively, characterizing  the non-integrable Ising model \eqref{NI}.}
        \label{fig:nine}
\end{figure}

Figure~\ref{fig:nine} shows the complexity as a function of the input amplitude for the Ising systems~\eqref{T} and \eqref{NI}. The amplitude-aware permutation entropy ranks the cases in the same order of complexity as the size of the minimal viable latent space for an auto-encoder network.

\section{Conclusion}

In this study, we demonstrated the application of Temporal Convolutional Networks (TCNs) for input-output time series analysis in the context of optics, specifically focusing on quantum spin systems subjected to an external time-dependent magnetic field. We have shown that given a time series input-output dataset, there exists a unique minimum TCN architecture that is able to accurately model the input-output mapping. In particular we propose that the minimum viable dimension of the latent space can be used to quantify the complexity of the underlying system. 

Our findings show that systems in non-chaotic regimes can be modeled with much simpler TCN architectures compared to chaotic regimes, reflecting inherent complexity differences.  Additionally, we compared our TCN-based complexity measure with amplitude-aware permutation entropy and found that both measures align, ranking the respective mappings in the same order of complexity. We note, however, that the definition of entropy in time series data remains an active area of research; there is no universally accepted method due to the diverse nature of time series characteristics and the varied objectives of entropy analysis~\cite{complexities1,complexities2,complexities3,permutation1,complexities5,complexities6,complexities7,complexities8,complexities9,complexities10}. The proposed latent-space measure not only contributes to the existing repertoire of complexity measures but is argued to be more robust. This robustness stems from the efficiency of convolutional networks in feature extraction. We suggest that TCNs capture the complexity of time series data more effectively than alternative methods.

In our future work, we aim to address issues identified with our current approach. One primary concern is the lack of smoothness observed in the predicted output. To address this, we will investigate potential causes of this phenomenon and explore whether post-processing can mitigate it. Additionally, we will explore algorithms beyond TCN to improve model robustness and efficiency, aiming for better performance and reduced computational time.

\emph{Code availability:} All codes used in this study can be found in~\cite{Generative-ML-Quantum-System2025}.

\acknowledgments

This work has been supported by Army Research Office (ARO) (grant W911NF-23-1-0288; program manager Dr.~James Joseph). The views and conclusions contained in this document are those of the authors and should not be interpreted as representing the official policies, either expressed or implied, of ARO or the U.S. Government. The U.S. Government is authorized to reproduce and distribute reprints for Government purposes notwithstanding any copyright notation herein.

\appendix
\section{Traditional CNN Autoencoder} \label{appendix1}

Let us explore why the traditional Convolutional Neural Network (CNN)-based autoencoder model often falls short when applied to time-series analysis. To understand this limitation, it is essential to highlight the key differences between CNNs and Temporal Convolutional Networks (TCNs). At a fundamental level, CNNs are specifically designed for spatial data, such as images, where patterns and features are extracted from fixed regions in two-dimensional space. On the other hand, TCNs are optimized for sequential data, making them well-suited for time-series analysis. In TCN, this is achieved by incorporating mechanisms like dilated convolutions and causal padding, which enable them to effectively capture temporal dependencies across different time steps, a task that CNNs struggle with due to their lack of temporal awareness.

Here, we focus on \textbf{Case 1} from Sec.~\ref{Section 2} to demonstrate model performance on this simple dataset $\mathcal{D}(A=1)$. As the amplitude increases, the dataset complexity grows, but even with this basic case, the CNN-based model performs poorly compared to the TCN-based model implemented in this paper.

\begin{figure}
        \centering
        \includegraphics[width=\linewidth]{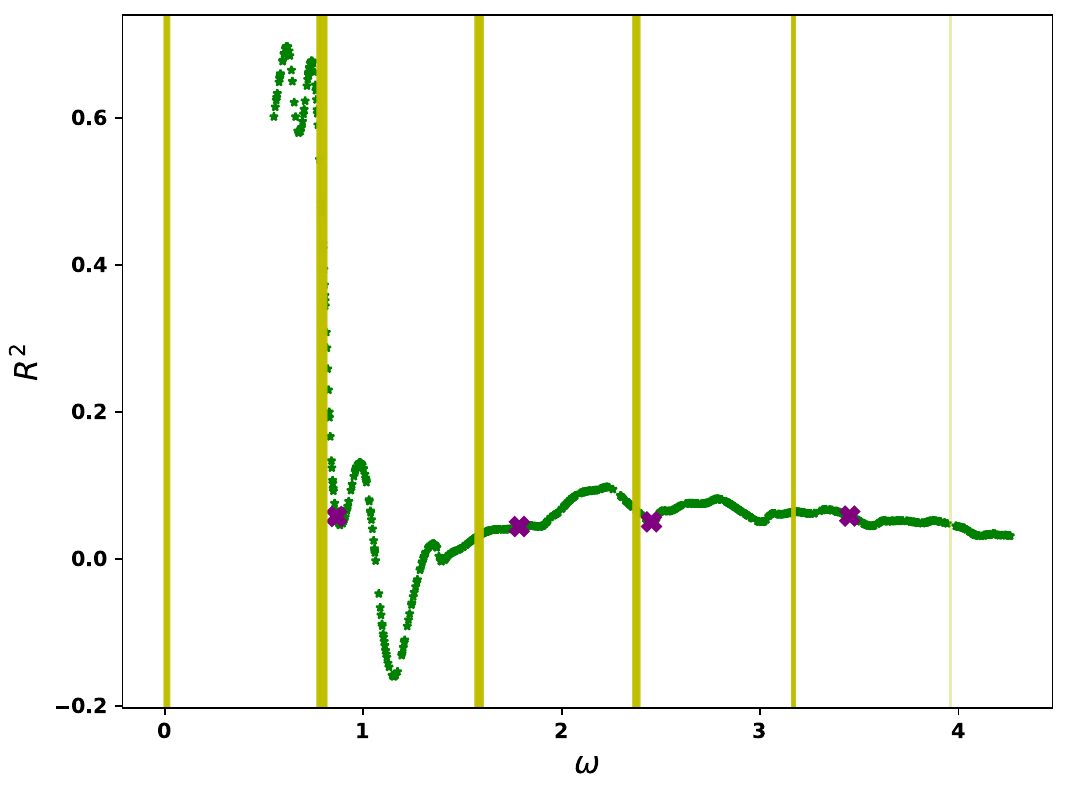}
        \caption{Plot shows the $R^{2}$ value vs frequency $\omega$ for the test dataset. Here we have used the CNN autoencoder model architecture: $(5-5-3)$.}
        \label{fig:CNN_V1}
\end{figure}

Case (a): Architecture : $(5-5-3)$, Number of parameters $= 294$. In this case, we used the same architecture as in \textbf{Case 1} of the TCN model, but replaced the TCN layers with CNN layers. The resulting model demonstrates significantly lower performance compared to its TCN counterpart, as shown in Fig.~\ref{fig:CNN_V1}.  

\begin{figure}
    \centering
    \includegraphics[width=\linewidth]{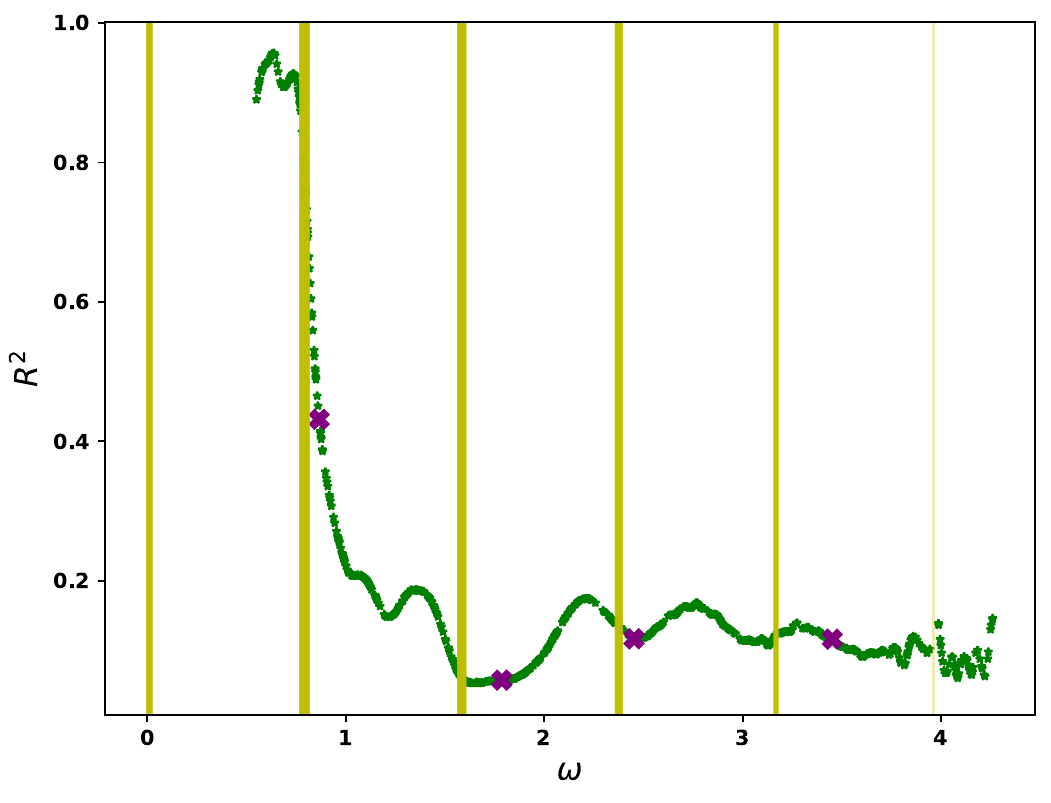}
    \caption{Plot shows the $R^{2}$ value vs frequency $\omega$ for the test dataset. Here we have used the CNN autoencoder model architecture: $(32-16-8-6)$.}
    \label{fig:CNN_V2}
\end{figure}

Case (b): Architecture : $(32-16-8-6)$, Number of parameters $= 4,439$. We implemented a CNN-based autoencoder model with a comparable number of parameters to the TCN-based model to ensure a fair comparison of performance. The $R^2$ value shows slight improvement over Case (a), but it remains significantly lower than the TCN model. The performance is illustrated in Fig. \ref{fig:CNN_V2}.  

\section{Minimum Stable Architecture}\label{appendix2}

We will elucidate the concept of a minimum stable architecture. To accomplish this, we will examine \textbf{Case 1} from Sec.~\ref{Section 2} and analyze it systematically across various architectures. Specifically, we will demonstrate that any architecture with fewer parameters than the minimum stable architecture (5-5-3), which comprises 2082 parameters, not only exhibits lower performance but also lacks stability.

Table \ref{table_5-5-3} summarizes the performance of the (5-5-3) architecture across 10 different training runs. In other words, it shows the performance of 10 different models built using the same architecture. The first column lists the model numbers, while the second column indicates the percentage (rounded to the nearest integer) of values of \( \{ R_{i}^2 \mid i \in \text{TestSet} \} \) greater than 0.98. Similarly, the third column shows the percentage of values exceeding 0.95, and the fourth column captures the percentage of values above 0.90.

\begin{table}[H]
\centering
\begin{tabular}{|c|c|c|c|}
\hline
\textbf{Model No.} & \textbf{Above 0.98} & \textbf{Above 0.95} & \textbf{Above 0.90} \\ \hline \hline
1 & 98 & 99 & 100 \\ 
2 & 96 & 99 & 100 \\ 
3 & 97 & 100 & 100 \\ 
4 & 82 & 98 & 99 \\ 
5 & 93 & 99 & 100 \\ 
6 & 94 & 100 & 100 \\ 
7 & 94 & 100 & 100 \\ 
8 & 89 & 99 & 100 \\ 
9 & 89 & 99 & 100 \\ 
10 & 98 & 100 & 100 \\ \hline
\textbf{Average} & \textbf{93} & \textbf{99} & \textbf{100} \\ \hline
\end{tabular}
\caption{Performance of (5-5-3) Architecture with 2082 trainable parameters}
\label{table_5-5-3}
\end{table}

Table \ref{table_5-5-2} summarizes the performance of the (5-5-2) architecture in 10 different training runs. Note that we have reduced the latent dimension, which results in a lower number of trainable parameters: 1960, which is just 122 parameters fewer than in the previous model. Although we should not expect a significant change in behavior, we observe two key outcomes. First, the average percentage of values of \( \{ R_{i}^2  \mid i \in \text{TestSet} \} \) greater than 0.98 decreases. Second, in one such run, this value drops to 77, marking the first instance of instability. In this context, instability is defined as the emergence of significant deviations in performance metrics, such as the drop to 77\% for \( \{ R_{i}^2 \mid i \in \text{TestSet} \} \) above 0.98, across specific runs. This indicates reduced robustness and reliability, likely due to the decreased number of trainable parameters in the modified architecture.
\begin{table}
\centering
\begin{tabular}{|c|c|c|c|}
\hline
\textbf{Model No.} & \textbf{Above 0.98} & \textbf{Above 0.95} & \textbf{Above 0.90} \\ \hline \hline
1 & 89 & 98 & 99 \\ 
2 & 89 & 99 & 99 \\ 
3 & 77 & 99 & 100 \\ 
4 & 95 & 100 & 100 \\ 
5 & 93 & 99 & 100 \\ 
6 & 80 & 91 & 99 \\ 
7 & 91 & 100 & 100 \\ 
8 & 83 & 99 & 100 \\ 
9 & 80 & 91 & 98 \\ 
10 & 85 & 99 & 99 \\ \hline
\textbf{Average} & 86.2 & 97.6 & 99.4 \\ \hline
\end{tabular}
\caption{Performance of (5-5-2) Architecture with 1960 trainable parameters}
\label{table_5-5-2}

\end{table}
In Table \ref{table_4-4-3}, we present the performance summary of the (4-4-3) architecture. This configuration reduces the number of trainable parameters to 1,433, representing a decrease of 649 parameters compared to the minimal stable model. 
\begin{table} 
\centering
\begin{tabular}{|c|c|c|c|}
\hline
\textbf{Model No.} & \textbf{Above 0.98} & \textbf{Above 0.95} & \textbf{Above 0.90} \\ \hline \hline
1 & 98 & 99 & 100 \\ 
2 & 87 & 99 & 99 \\ 
3 & 74 & 94 & 100 \\ 
4 & 94 & 100 & 100 \\ 
5 & 85 & 100 & 100 \\ 
6 & 82 & 98 & 100 \\ 
7 & 94 & 99 & 100 \\ 
8 & 78 & 92 & 96 \\ 
9 & 79 & 97 & 100 \\ 
10 & 84 & 99 & 99 \\ \hline
\textbf{Average} & 85.5 & 97.7 & 99.4 \\ \hline
\end{tabular}
\caption{Performance of (4-4-3) Architecture with 1433 trainable parameters}
\label{table_4-4-3}
\end{table}

The average percentage of values of \( \{ R_{i}^2  \mid i \in \text{TestSet} \} \) greater than 0.98 drops to 85.5\%, compared to 86.2\% in the (5-5-2) architecture and 93\% in the (5-5-3) architecture. Multiple runs show percentages below 80\%, with a minimum value of 74\%, highlighting the growing instability as the parameters are reduced.
In Table \ref{table_3-3-2}, we present the performance summary of the (3-3-2) architecture. This configuration reduces the number of trainable parameters to 810, representing a significant decrease compared to the (5-5-3) architecture with 2082 parameters. Specifically, the reduction amounts to 1272 parameters (61.1\%) from the (5-5-3) architecture.
\begin{table}
\centering
\begin{tabular}{|c|c|c|c|}
\hline
\textbf{Model No.} & \textbf{Above 0.98} & \textbf{Above 0.95} & \textbf{Above 0.90} \\ \hline \hline
1 & 61 & 77 & 90 \\ 
2 & 79 & 93 & 99 \\ 
3 & 76 & 93 & 100 \\ 
4 & 63 & 81 & 94 \\ 
5 & 80 & 93 & 97 \\ 
6 & 76 & 81 & 99 \\ 
7 & 78 & 95 & 99 \\ 
8 & 89 & 98 & 100 \\ 
9 & 82 & 99 & 100 \\ 
10 & 80 & 96 & 99 \\ \hline
\textbf{Average} & 85.5 & 97.7 & 99.4 \\ \hline
\end{tabular}
\caption{Performance of (3-3-2) Architecture with 810 trainable parameters}
\label{table_3-3-2}
\end{table}

\begin{figure}[H]
        \centering
        \includegraphics[width=\linewidth]{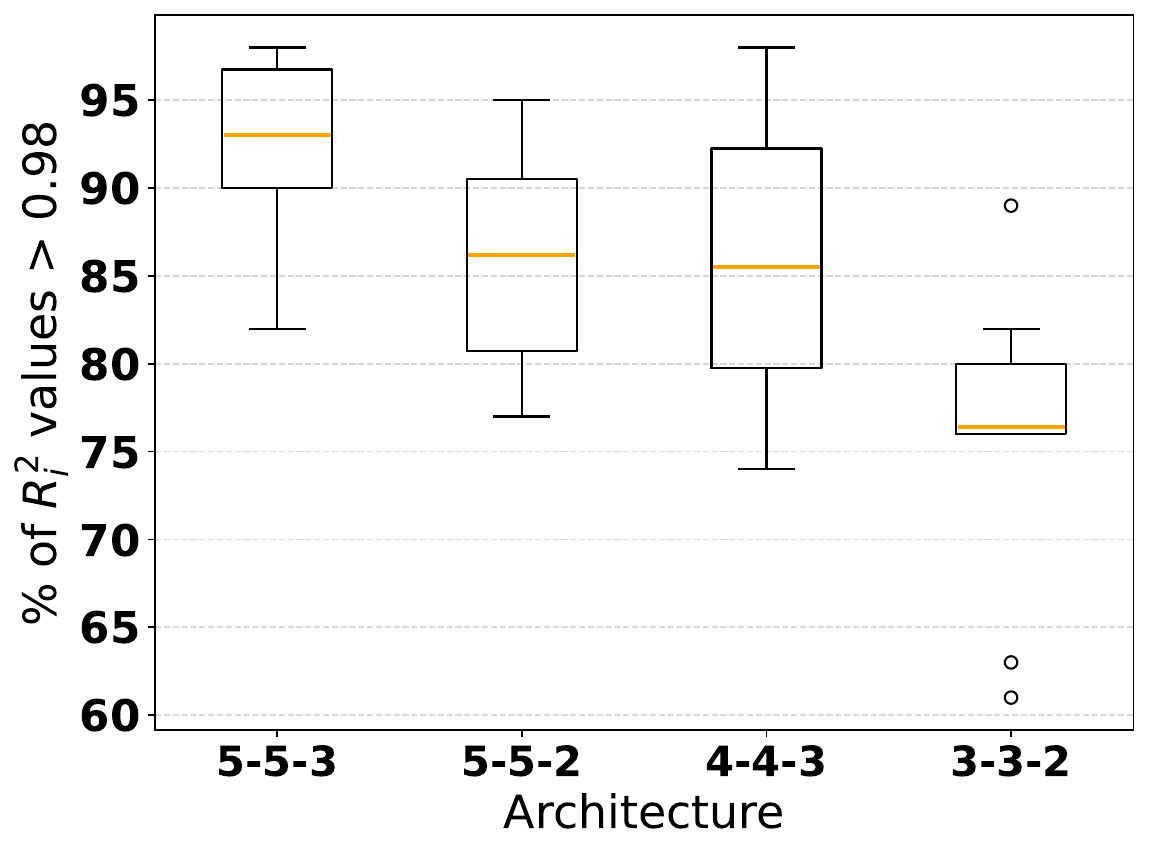}
        \caption{This box plot visually summarizes the performance of different architectures (5-5-3, 5-5-2, 4-4-3, and 3-3-2) evaluated on \( \{ R_i^2 \mid i \in \text{TestSet} \} \) values exceeding 0.98 across ten training runs. The orange horizontal line within each box represents the mean performance, providing a clear depiction of the average behavior across runs, as also depicted in the accompanying tables \ref{table_5-5-3},\ref{table_5-5-2},\ref{table_4-4-3} and \ref{table_3-3-2}. The 5-5-3 architecture demonstrates the highest mean and least variability, confirming its stability and superior performance. Conversely, as the number of trainable parameters decreases (e.g., 4-4-3 and 3-3-2), the mean performance drops, and variability increases, indicating growing instability.
}\label{Figboxplot_case1}
\end{figure}

Figure~\ref{Figboxplot_case1} summarizes the analysis.
It is clear that the instability has become more prominent.

\section{Minimum Stable Architecture: Comparison of VAE-TCN versus TCN}\label{appendix3}

\begin{figure}
        \centering
        \includegraphics[width=\linewidth]{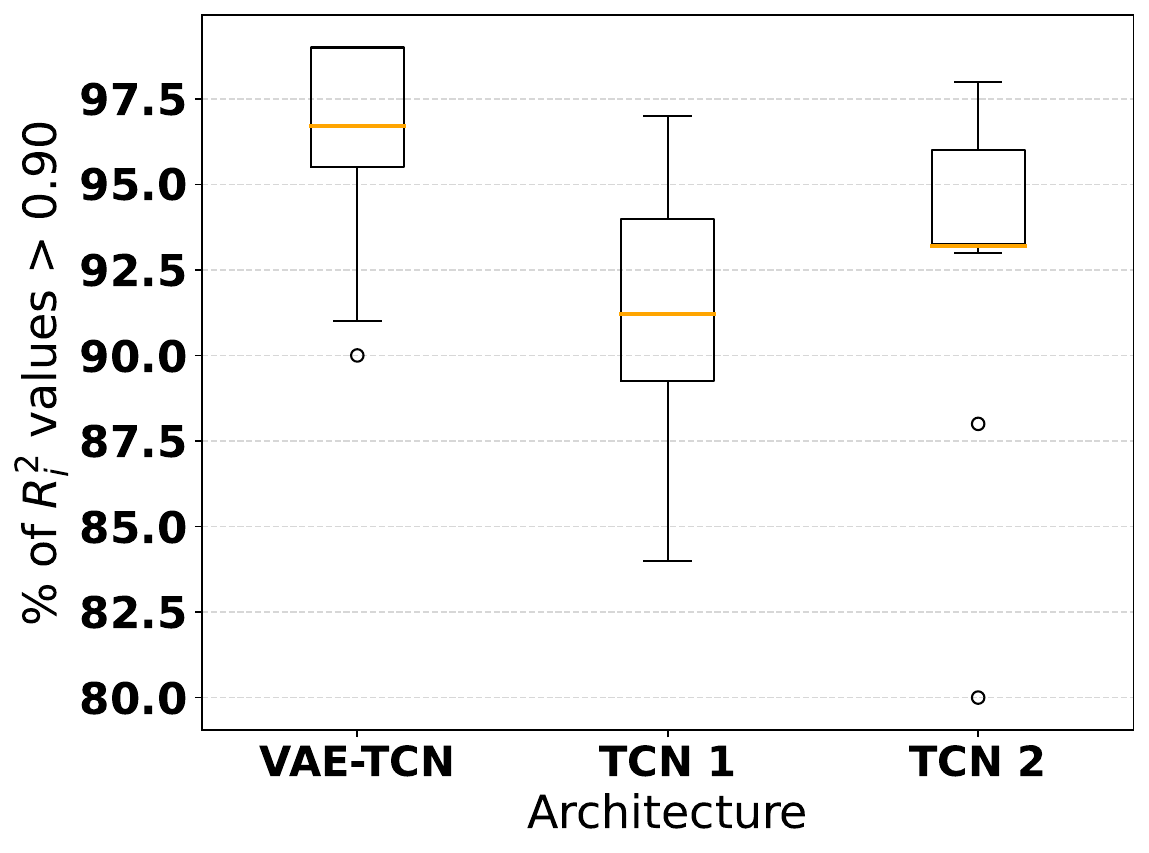}

        \caption{This box plot compares the performance of three models: VAE-TCN, TCN 1, and TCN 2, evaluated on \( \{ R_i^2 \mid i \in \text{TestSet} \} \) values exceeding 0.90 across ten training runs. The orange horizontal line within each box represents the mean performance, providing a clear depiction of the average behavior across runs, as summarized in Table \ref{table_vae_tcn}. The VAE-TCN model achieves the highest mean performance, effectively capturing intricate features. While TCN 1, the minimal stable architecture, is stable, TCN 2, despite a similar parameter count to VAE-TCN (31,243 vs. 32,127), performs marginally lower, highlighting VAE-TCN's superiority.
}

        \label{VAETCN}
\end{figure}

In Sec.~\ref{subD}, we demonstrate how the VAE-TCN model excels at capturing subtle and intricate features, resulting in a superior fit for \textbf{Case 2}  from Sec.~\ref{Section 2}. To provide a thorough analysis, we compare three models: the (12-12-10-10) VAE-TCN model, the (12-12-10-10) TCN model (TCN 1), and the (18-18-13-12) TCN model (TCN 2). The inclusion of the third model in this comparison is crucial. The results emphasize that even when the TCN model with the architecture (18-18-13-12) and a parameter count of 31,243 closely aligns with the VAE-TCN model's parameter count of 32,127, its performance remains comparatively lower. This discrepancy can be better understood by revisiting the key insights discussed in Sec.~\ref{subD}. Additionally, it is important to note that the TCN 1 model demonstrates stable performance despite having a parameter count of 16,127, which is significantly lower compared to the other models. The comparative performance of the three models is visually summarized in Fig.~\ref{VAETCN}.

\begin{table}
\centering
\begin{tabular}{ |p{1.8cm}|p{1.8cm}|p{1cm}|p{1cm}| }
 \hline
 \textbf{Model No.} & \textbf{VAE-TCN} & \textbf{TCN 1} & \textbf{TCN 2}  \\ 
 \hline \hline
 1 & 99 & 85 & 93 \\ 
 2 & 99 & 95 & 96 \\ 
 3 & 99 & 90 & 95 \\ 
 4 & 91 & 92 & 97 \\ 
 5 & 90 & 97 & 95 \\ 
 6 & 97 & 94 & 88 \\ 
 7 & 99 & 89 & 96 \\ 
 8 & 99 & 92 & 98 \\ 
 9 & 95 & 84 & 94 \\ 
 10 & 99 & 94 & 80 \\ 
 \hline
 \textbf{Average} & 96.7 & 91.2 & 93.2 \\ 
 \hline
\end{tabular}
\caption{Performance of different models.}
\label{table_vae_tcn}
\end{table}

\bibliography{References.bib} 

\end{document}